\documentclass[twocolumn,numberedappendix,twocolappendix,iop,apj]{aastex7}
\usepackage{color,graphicx,natbib}
\usepackage{graphicx}
\usepackage{amsmath, amssymb, bm}
\usepackage{amssymb}
\usepackage{arydshln}
\usepackage{bm}
\usepackage[toc,page]{appendix}
\usepackage[T1]{fontenc}
\usepackage{xspace}
\usepackage{booktabs}
\usepackage[toc,page]{appendix}
\usepackage{float}
\usepackage{caption}  
\usepackage{subcaption}
\usepackage{hyperref}
\usepackage{tikz}
\usetikzlibrary{arrows.meta, positioning, shapes.geometric}
\hypersetup{
    colorlinks=True,
    citecolor=orange,
    linkcolor=black,
    }

\shorttitle{A Systematic Framework for Validating Global MVT of GRBs}
\shortauthors{Bala et al. 2025}

\begin{document}

\title{Fastest or Significant: A Systematic Framework for Validating Global Minimum Variability Timescale Measurements of Gamma-ray Bursts}

\correspondingauthor{Suman Bala}
\email{sumanbala2210@gmail.com}

\author[0000-0002-6657-9022]{S.~Bala}
\affiliation{Science and Technology Institute, Universities Space Research Association, Huntsville, AL 35805, USA}
\email{sbala@usra.edu}

\author[0000-0002-2149-9846]{P.~Veres}
\affiliation{Department of Space Science, University of Alabama in Huntsville, Huntsville, AL 35899, USA}
\affiliation{Center for Space Plasma and Aeronomic Research, University of Alabama in Huntsville, Huntsville, AL 35899, USA}
\email{pv0004@uah.edu}

\author[0000-0002-0587-7042]{A.~Goldstein}
\affiliation{Science and Technology Institute, Universities Space Research Association, Huntsville, AL 35805, USA}
\email{AGoldstein@usra.edu}

\author[0009-0009-2018-9457]{R.~Sonawane}
\affiliation{School of Physics, Indian Institute of Science Education and Research Thiruvananthapuram, Thiruvananthapuram, 695551, India}
\affiliation{Center for High Performance Computing, Indian Institute of Science Education and Research Thiruvananthapuram, Thiruvananthapuram, 695551, India}
\email{rushikesh23@iisertvm.ac.in}

\author[0009-0008-4455-7931]{R.~Samanta}
\affiliation{School of Physics, Indian Institute of Science Education and Research Thiruvananthapuram, Thiruvananthapuram, 695551, India}
\email{rupam24@iisertvm.ac.in}

\author[0000-0003-3220-7543]{S.~Iyyani}
\affiliation{School of Physics, Indian Institute of Science Education and Research Thiruvananthapuram, Thiruvananthapuram, 695551, India}
\affiliation{Center for High Performance Computing, Indian Institute of Science Education and Research Thiruvananthapuram, Thiruvananthapuram, 695551, India}
\email{shabnam@iisertvm.ac.in}



\begin{abstract}
The minimum variability timescale (MVT) is a key observable used to probe the central engines of Gamma--Ray Bursts (GRBs) by constraining the emission region size and the outflow Lorentz factor. However, its interpretation is often ambiguous: statistical noise and analysis choices can bias measurements, making it difficult to distinguish genuine source variability from artifacts. Here we perform a comprehensive suite of simulations to establish a quantitative framework for validating Haar--based MVT measurements. We show that in multi--component light curves, the MVT returns the most statistically significant structure in the interval, which is not necessarily the fastest intrinsic timescale, and can therefore converge to intermediate values. Reliability is found to depend jointly on the MVT value and its signal--to--noise ratio ($\mathrm{SNR}_{\mathrm{MVT}}$), with shorter intrinsic timescales requiring proportionally higher $\mathrm{SNR}_{\mathrm{MVT}}$ to be resolved.

We use this relation to define an empirical MVT Validation Curve, and provide a practical workflow to classify measurements as robust detections or upper limits. Applying this procedure to a sample of \textit{Fermi}--GBM bursts shows that several published MVT values are better interpreted as upper limits. These results provide a path toward standardizing MVT analyses and highlight the caution required when inferring physical constraints from a single MVT measurement in complex events.
\end{abstract}

\keywords{gamma-ray burst: general --- variability time scales, Gamma-ray Burst Monitor (GBM)}

\section{Introduction}
\label{sec:intro}

Gamma-ray bursts (GRBs) are the most luminous electromagnetic transients in the Universe, produced at cosmological distances during catastrophic compact-object events. They are broadly classified into two groups: long-duration GRBs (T$_{90} > 2$ s), generally associated with the core-collapse of massive stars \citep{Woosley93, 1998ApJ...494L..45P, macfadyen1999collapsars, 2006ARA&A..44..507W, Yoon05}, and short-duration GRBs (T$_{90} < 2$ s), linked to the merger of compact objects \citep{Eichler89, 1992ApJ...392L...9D, kmf93, 2006ApJ...638..354B, Tanvir2013, Goldstein17, LIGO-Fermi17}. 
While parameters such as spectral lags \citep{nor96} and hardness ratios \citep{Paciesas99cat, Bhat_16, Kienlin2020} have been used to distinguish these populations, overlapping properties can complicate this classification.
The lGRBs are usually associated with Supernovae Ic-BL \citep{Hjorth2003} and sGRBs are associated with kilonovae \citep{Tanvir2013, Yang_2015, abb17, Levan_2023, Yang2024}. A key observable that may offer a more robust criterion is the Minimum Variability Timescale (MVT), the shortest statistically significant timescale on which the source flux varies. This is particularly relevant given recent observations of long-duration bursts like GRB 211211A \citep{Veres_2023, Rastinejad22,Gompertz23,Yang22,Troja22} and GRB 230307A \citep{dalessi_2025_grb_230307a, Du24a, Dai24}  which, despite their duration, are associated with kilonovae characteristic of merger events.

The physical significance of the MVT stems from its causal relation to the size of the emission region, $R$, and the bulk Lorentz factor of the outflow, $\Gamma$, via the relation $\delta t_{\min} \lesssim R / (c\,\Gamma^{2})$. An accurate measurement of the MVT can therefore provide powerful constraints on the jet dynamics, help resolve the "compactness problem," and offer insights into the scale of the central engine itself \citep{luh08, rmm94}. Different emission models, such as internal shocks \citep{rmm94, lei07} or photospheric models \citep{ryde04}, also predict different variability characteristics, making the MVT a vital tool for constraining GRB progenitor and emission theories.

Despite its diagnostic power, reliably estimating the MVT from observational data remains difficult. GRB light curves are affected by Poisson noise that can both mimic and mask fast structure, and different estimators, including pulse fitting \citep{Norris96,Norris05,bhat12,2025arXiv250808995M}, wavelet analysis \citep{mac13,Golkhou_2014,Golkhou15,Vianello18}, and measuring the full–width at half–maximum of the shortest significant pulse \citep{Camisasca23,2025arXiv250808995M}, sometimes yield discrepant values for the same GRB. Although these methods have been used in many studies, a systematic calibration across the relevant parameter space — particularly for very short intrinsic timescales ($\sim$few ms) and at high signal strengths — is still lacking. It therefore remains unclear whether existing techniques can always recover the fastest true timescale under these conditions. This concern is reinforced by recent observations of extreme variability, such as sub–millisecond features in GRB 200415A \citep{2021Natur.589..207R} and complex structures in very bright events such as GRB 221009A \citep{lesage2023fermi,burns2024fermigbmteamanalysisravasio} and GRB 230307A, which highlight the need for a quantitative framework to determine when an observed MVT reflects genuine variability rather than statistical artifacts.

GB14 \citep{Golkhou_2014} demonstrated that the Haar wavelet transform can be used to estimate short variability timescales in GRB prompt emission by identifying the smallest dyadic scale at which the wavelet variance rises above the Poisson noise floor. This provides a physically interpretable, algorithmic definition of the MVT that does not rely on assuming a parametric pulse model. For this reason, and because the Haar estimator responds directly to statistically significant structure rather than to individual fitted components, we adopt the same Haar–based definition here. What GB14 did not address, however, is under what observational conditions this timescale can be \emph{reliably} measured. In particular, GB14 did not explore how the method behaves in the extreme regions of parameter space — where very short intrinsic timescales, very high SNR, and complex / multi–component structures occur simultaneously.
 In this paper we extend that earlier work by quantifying the behavior of the Haar–based MVT across a controlled suite of simulations where the true intrinsic variability timescale is known, and by developing an empirical framework that can be used to validate MVT measurements extracted from real data.

Section~\ref{sec:simulations} describes the simulation methodology. Section~\ref{sec:results} presents the results of the analysis. Section~\ref{sec:real_grbs} applies the framework to a sample of real GRBs, and Section~\ref{sec:discussion} summarizes the key findings and conclusions.


\section{Simulations}
\label{sec:simulations}

To evaluate and understand the performance of the Minimum Variability Timescale (MVT) analysis, we developed a simulation framework that systematically explores a multi-dimensional parameter space. In each simulation campaign, we varied key pulse parameters while keeping others, such as the background rate, fixed. The framework constructs a grid of all possible parameter combinations, with each grid point representing a distinct physical scenario. For \textbf{every scenario, we generated 300 independent realizations} of time-tagged event (TTE) data based on analytical pulse models. The resulting ensemble yields a distribution of MVT values, enabling us to quantify statistical fluctuations and assess the stability of the analysis across different physical conditions.

\subsection{Lightcurve Simulations}
\label{subsec:generation}

For idealized tests, we generated raw photon arrival times without instrument-specific effects. This process uses an inverse--transform sampling method on the total (source + background) rate function, $R(t)$, defined over a fine, adaptive time grid. For these simulations, a constant \textbf{background rate of 1000 counts~s$^{-1}$} was used. The variable source strengths were then tested against this, here the \textbf{`peak amplitude'} is defined as the peak source rate divided by the background rate. The resolution of the time grid is determined by the narrowest temporal feature of the source pulse (e.g., rise time or standard deviation) and is clipped to a range of 0.1~$\mu$s to 1~ms to ensure both accuracy and computational stability. The total number of events for a given realization, $N_{\text{tot}}$, is drawn from a Poisson distribution based on the total expected counts, $\int_{\mathrm{t}_{\text{start}}}^{\mathrm{t}_{\text{stop}}} \mathrm{R(t') dt'}$. 

\begin{figure}[ht]
    \centering
    \includegraphics[width=0.45\textwidth]{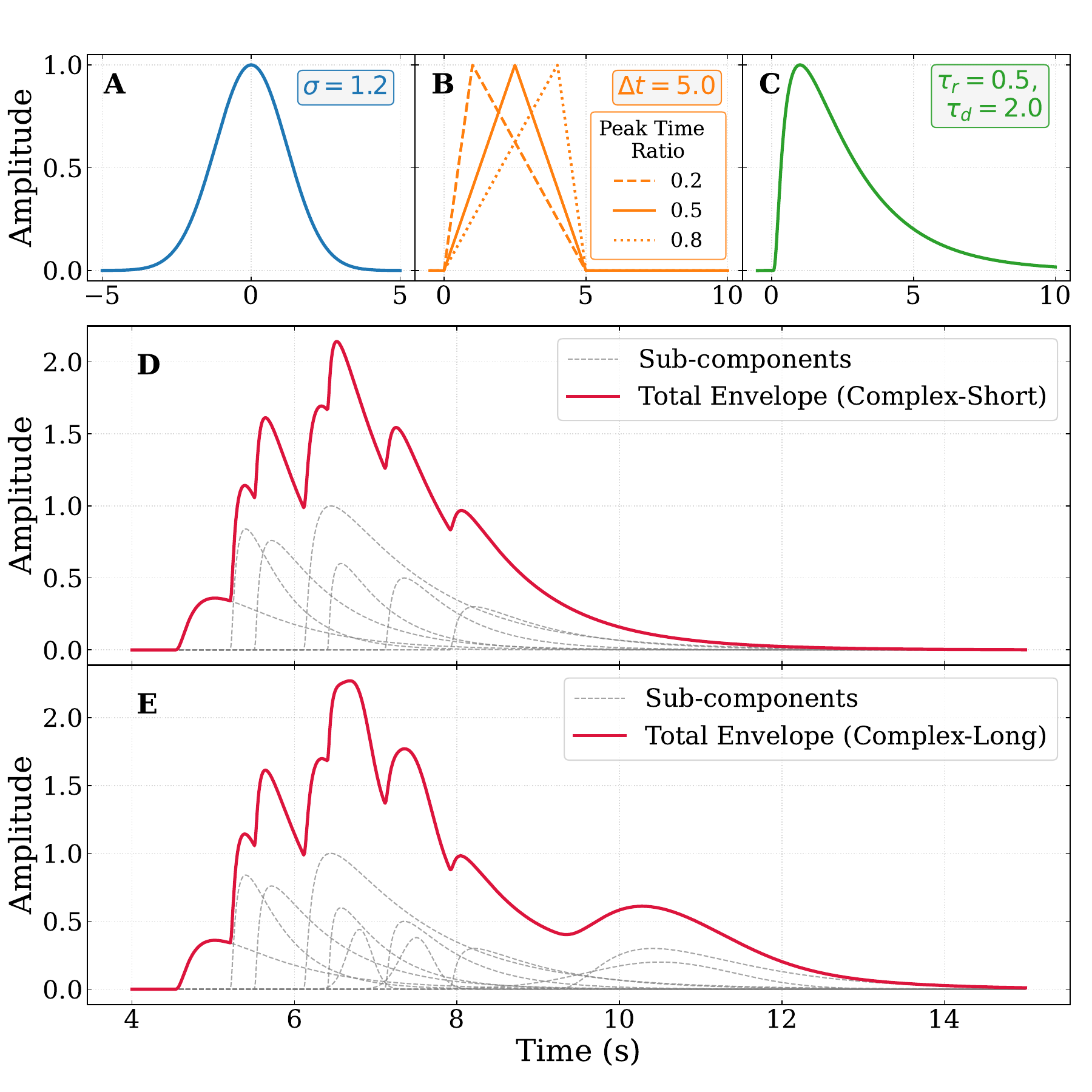}
    \caption{\small Representative pulse profiles and complex light curve templates used to calibrate and validate the MVT--$\mathrm{SNR}_{\mathrm{MVT}}$ relation. (A) Symmetric Gaussian profile defined by standard deviation $\sigma$. (B) Triangular pulses with a fixed duration $\Delta t = 5.0$ and varying peak time ratios. (C) Norris profile characterized by rise and decay timescales $\tau_r$ and $\tau_d$. (D, E) Complex, multi-component light curve templates constructed by summing individual sub-components.}
\label{fig:pulse_templates}
\end{figure}

\begin{table}[ht]
\centering
\caption{Component Definitions for the \textit{Complex lightcurve} Templates. This table details the 11 fixed components that constitute the underlying template emission. Additional ``feature'' pulses (Gaussian in this work) are then added to this template during the analysis phase. Parameters for Norris pulses are ($\mathrm{t}_{\text{start}}$, $\tau_{\text{rise}}$, $\tau_{\text{decay}}$), and for Gaussian pulses are ($\mu$, $\sigma$). All time parameters are in seconds. Representative figures are shown in Figure \ref{fig:complex_lc_example}.}
\label{tab:complex_pulse}
\begin{tabular}{l c c c c}
\midrule
\midrule
\textbf{Pulse} & \textbf{Rel. Amp.} & \textbf{Par 1} & \textbf{Par 2} & \textbf{Par 3} \\
\midrule
Norris   & 1.00 & 6.1  & 0.10 & 1.20 \\
Norris   & 0.84 & 5.2  & 0.08 & 0.50 \\
Norris   & 0.76 & 5.5  & 0.06 & 0.80 \\
Norris   & 0.60 & 6.4  & 0.05 & 0.60 \\
Norris   & 0.50 & 7.1  & 0.09 & 0.70 \\
Norris   & 0.30 & 7.9  & 0.10 & 1.00 \\
Norris   & 0.36 & 4.5  & 0.30 & 0.90 \\
Norris   & 0.30 & 9.0  & 2.00 & 1.00 \\
Gaussian & 0.44 & 6.8  & 0.15 & --   \\
Gaussian & 0.38 & 7.5  & 0.20 & --   \\
Gaussian & 0.20 & 10.5 & 0.90 & --   \\
\midrule
\end{tabular}
\end{table}
\subsection{Temporal Pulse Models and Simulation Windows}
\label{subsec:pulses}

The temporal evolution of the source flux was described using several distinct models. To ensure computational efficiency while capturing the full morphology of each pulse, the simulation time window ($\mathrm{t}_{\text{start}}$, $\mathrm{t}_{\text{stop}}$) and internal time grid resolution were calculated adaptively for each model based on its specific parameters:

\begin{itemize}
    \item \textbf{Gaussian}: A symmetric pulse defined by its standard deviation, $\sigma$. The simulation window was set to span $\pm 5\sigma$ around the pulse center, with additional buffer on each side proportional to the chosen grid resolution. The grid resolution was set to $\sigma / 10$.

    \item \textbf{Triangular (symmetric or asymmetric)}: A pulse defined by a linear rise and decay. Depending on the relative rise and fall durations, the pulse may be symmetric or asymmetric. For each simulation, the analysis window was set using the pulse start and stop times, with an additional resolution–scaled buffer applied at both ends. The grid resolution was chosen to be one–tenth of the shorter of the rise or fall durations.

    \item \textbf{Norris}: An empirical GRB pulse shape \citep{Norris2005}. The window was set to begin prior to the pulse onset and to end six decay timescales after the peak, ensuring that the full tail was included. The grid resolution was taken as one–tenth of the shorter of the rise or decay time.

    \item \textbf{Complex Light Curves}: 
    To represent complex, structured emission, two composite lightcurve templates were constructed from the summation of overlapping Norris and Gaussian pulses, with components detailed in Table~\ref{tab:complex_pulse}. The \textbf{`overall amplitude'} for these templates is defined as the peak source rate of the primary Norris pulse (with a relative amplitude of 1.0) divided by the background rate. To efficiently explore the parameter space, a two-stage process was used. First, a single, high-resolution template was generated. During analysis, a separate 'feature' pulse was then added to this template. The strength of this feature is defined by the `\textbf{relative peak amplitude' (RPA),} which is the ratio of the feature's peak amplitude to the `overall amplitude' of the template.
    \begin{itemize}
        \item \textbf{Complex-Long Lightcurve}: A template combining all 11 pulse components detailed in Table~\ref{tab:complex_pulse}. For these simulations, a fixed time window from 4.0~s to 15.0~s was used.
        \item \textbf{Complex-Short Lightcurve}: A template created by removing the four broadest pulse components to produce a shorter overall episode of variability. For these, a fixed time window from 4.0~s to 12.0~s was used.
    \end{itemize}
\end{itemize}

\subsection{MVT Calculation and Uncertainty Quantification}
\label{subsec:uncertainty}

We processed the simulated TTE data by generating binned light curves for each unique simulation condition, systematically varying key parameters such as the light curve \textbf{bin width (BW)}. Because a single MVT measurement from one light curve is insufficient to characterize the underlying variability timescale, we employ a Monte~Carlo approach to robustly quantify its statistical uncertainty.

For each fixed set of input parameters, we generate a large ensemble of $\mathrm{N}$ independent light curve realizations (typically \( \mathrm{N} = 300 \), as described in section \ref{sec:simulations}) and compute the MVT for each realization, resulting in a distribution of MVT values. We include in this distribution only those realizations that yield a valid measurement with non-zero uncertainty, ensuring the reliability of the statistical summary.

The \textbf{median (50th percentile)} of this distribution is reported as the final MVT. We use the median rather than the mean because the MVT distribution can be skewed and may contain outliers arising from stochastic fluctuations in individual realizations. The median provides a robust measure of the typical MVT, whereas the mean could be unduly influenced by extreme values. The associated asymmetric 68\% confidence interval is defined by \textbf{the 16th and 84th percentiles} of the distribution, reported as \((\mathrm{MVT}_{50} - \mathrm{MVT}_{16})\) for the lower bound and \((\mathrm{MVT}_{84} - \mathrm{MVT}_{50})\) for the upper bound. This percentile-based approach makes no assumption about the underlying distribution being Gaussian.

\section{Results}
\label{sec:results}
In this section, we present the results from a comprehensive suite of simulations designed to characterize the behavior of the Minimum Variability Timescale (MVT) algorithm under a variety of conditions. Our goal is to develop a systematic workflow to interpret MVT measurements from arbitrary light curves. We begin with a simple, idealized pulse shape and progressively introduce more complexity to test the universality of our findings.

\begin{figure}[ht]
    \centering
    \includegraphics[width=0.45\textwidth]{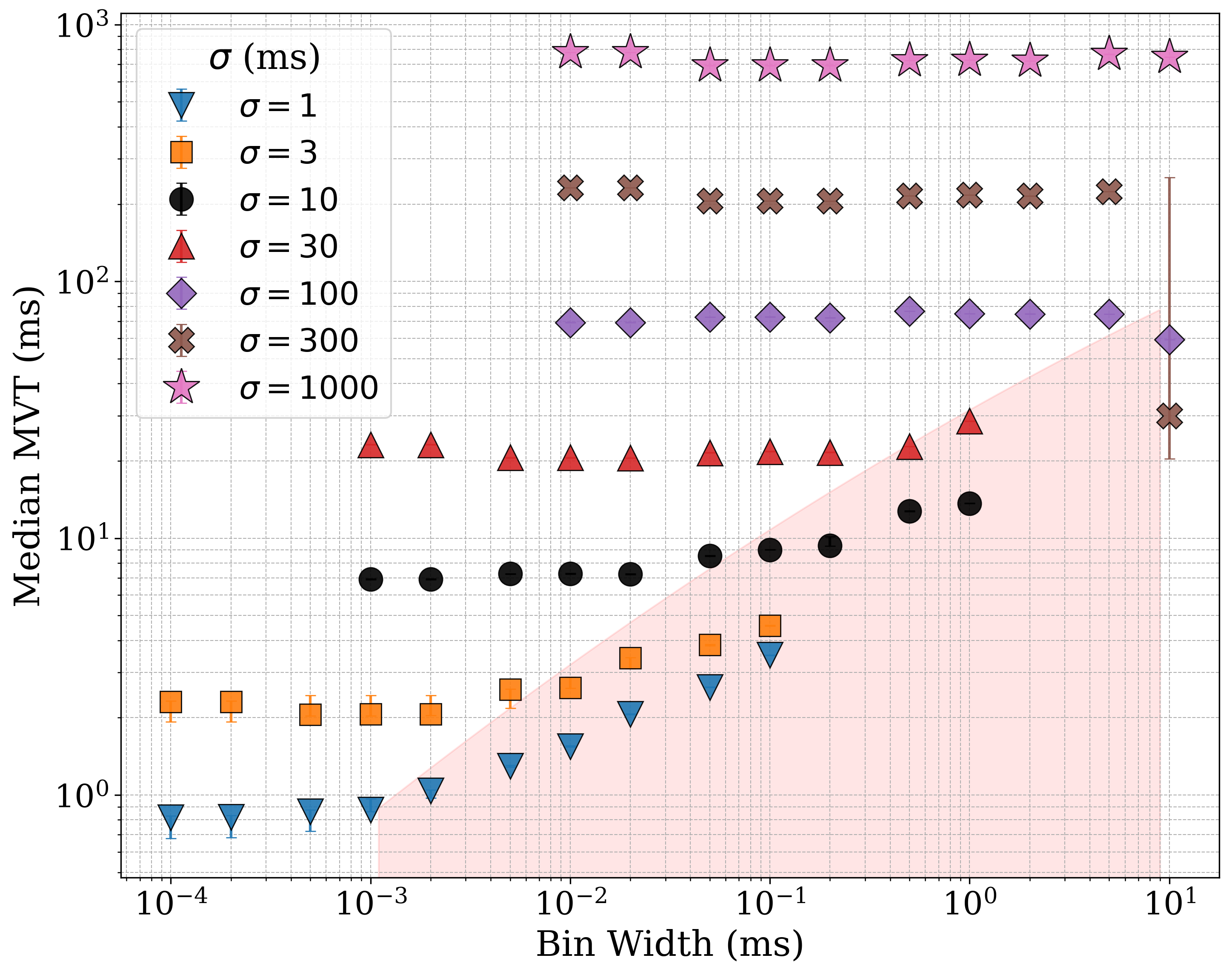}
    \caption{\small
        The median Minimum Variability Timescale (MVT) as a function of analysis bin width (BW) for high--SNR Gaussian pulses of varying intrinsic widths ($\sigma$). The plot shows two distinct regimes: a bin--limited regime (red region) at large BW where the MVT is systematically overestimated, and a source--dominated regime at small BW where the MVT converges to a stable plateau that reflects the true $\sigma$. This demonstrates that a BW significantly smaller than the timescale of interest is required for a reliable measurement.
        }

    \label{fig:gaussian_mvt_vs_bw}
\end{figure}

\subsection{The Ideal Case: Gaussian Pulses}
\label{subsec:gaussian}

We begin our investigation with one of the the simple pulse shape: a Gaussian profile. We simulated a large suite of light curves containing Gaussian pulses with intrinsic widths ($\sigma$) ranging from 1~ms to 1~s, with a constant background (1000~$\mathrm{counts\,s^{-1}}$) at various peak amplitudes.

Our initial test examines the role of the BW. To isolate this effect from statistical noise, we use a set of high-SNR simulations, with the results shown in Figure~\ref{fig:gaussian_mvt_vs_bw}. The plot shows two distinct regimes. When the BW is comparable to or larger than the intrinsic $\sigma$, the measurement is systematically overestimated and entirely dependent on the BW; we term this \textbf{the "bin-limited"} regime. Conversely, as the BW becomes significantly smaller than $\sigma$, the measured MVT becomes independent of the binning and converges to a stable plateau. In this \textbf{"source-dominated"} regime, the MVT accurately reflects the true $\sigma$. For broad pulses (e.g., $\sigma \ge 100$ ms), this plateau is quite wide, yielding a stable measurement even for BWs as large as 10 ms. This result, which is in agreement with the assumptions of GB14, demonstrates that a sufficiently small BW ($ \ll \sigma$) is a prerequisite for any meaningful MVT measurement.

\begin{figure}[ht]
    \centering
    \includegraphics[width=0.45\textwidth]{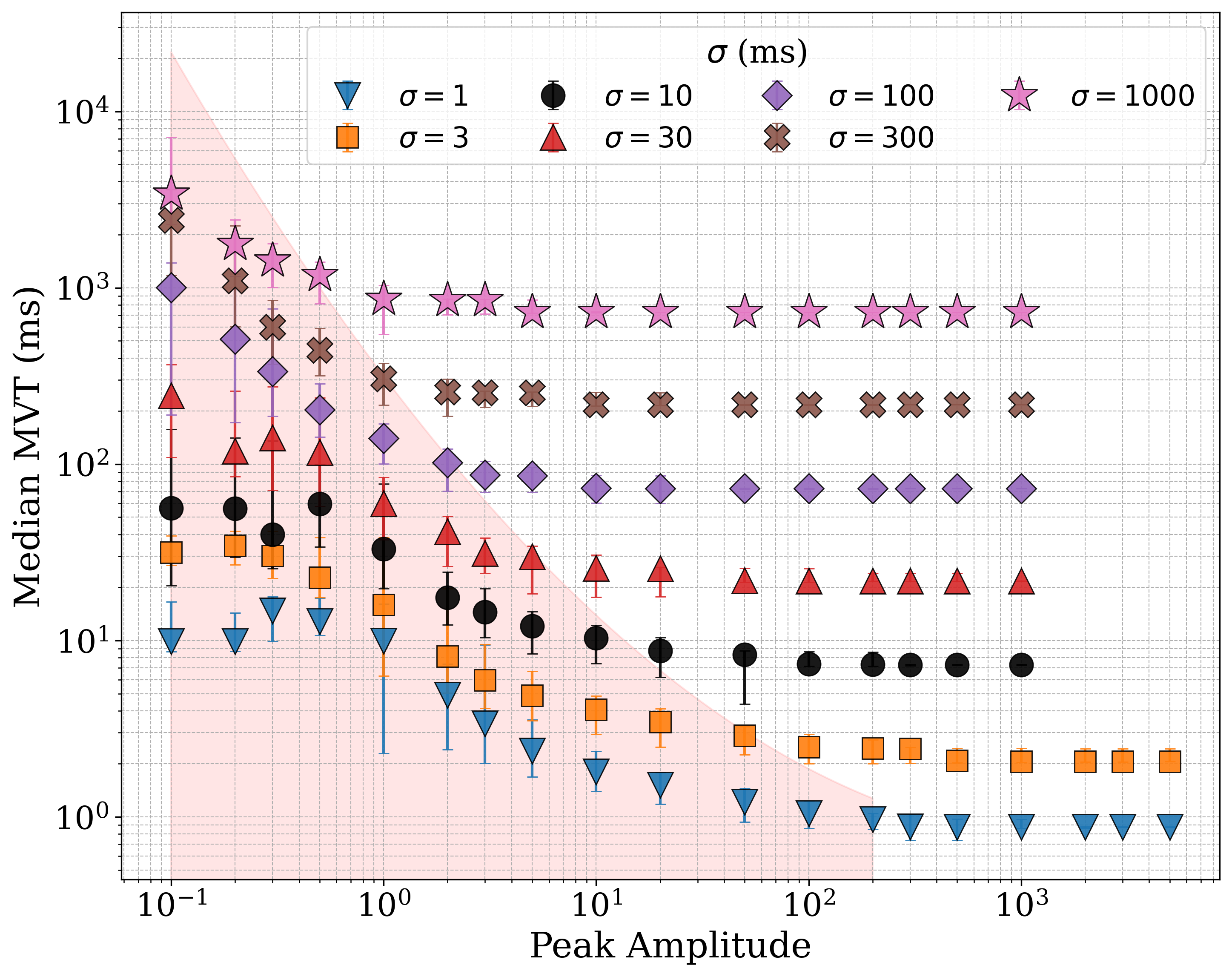}
    \caption{\small
    The median MVT as a function of peak count–rate amplitude for Gaussian pulses with varying intrinsic widths ($\sigma$). Each marker style represents a different $\sigma$, as shown in the legend. At low amplitudes (red region), the MVT is in the noise–dominated regime, yielding highly scattered and systematically overestimated values. As the amplitude increases, the MVT converges toward the true intrinsic width of the pulse.
    }

    \label{fig:gaussian_mvt_vs_amp}
\end{figure}

Next, we explore the effect of signal strength, using peak amplitude as a direct proxy. Figure~\ref{fig:gaussian_mvt_vs_amp} plots the measured MVT versus the pulse amplitude for several intrinsic widths ($\sigma$), analyzed at an appropriately small BW. At low amplitudes, the measurements are highly scattered and systematically overestimated, characterizing a noise-dominated regime. As the amplitude increases, the MVT for each $\sigma$ value converges to its true intrinsic width, confirming that a sufficient signal strength is necessary for an accurate measurement. This effect is particularly pronounced for pulses with smaller intrinsic widths.

\begin{figure}[ht]
    \centering
    \includegraphics[width=0.45\textwidth]{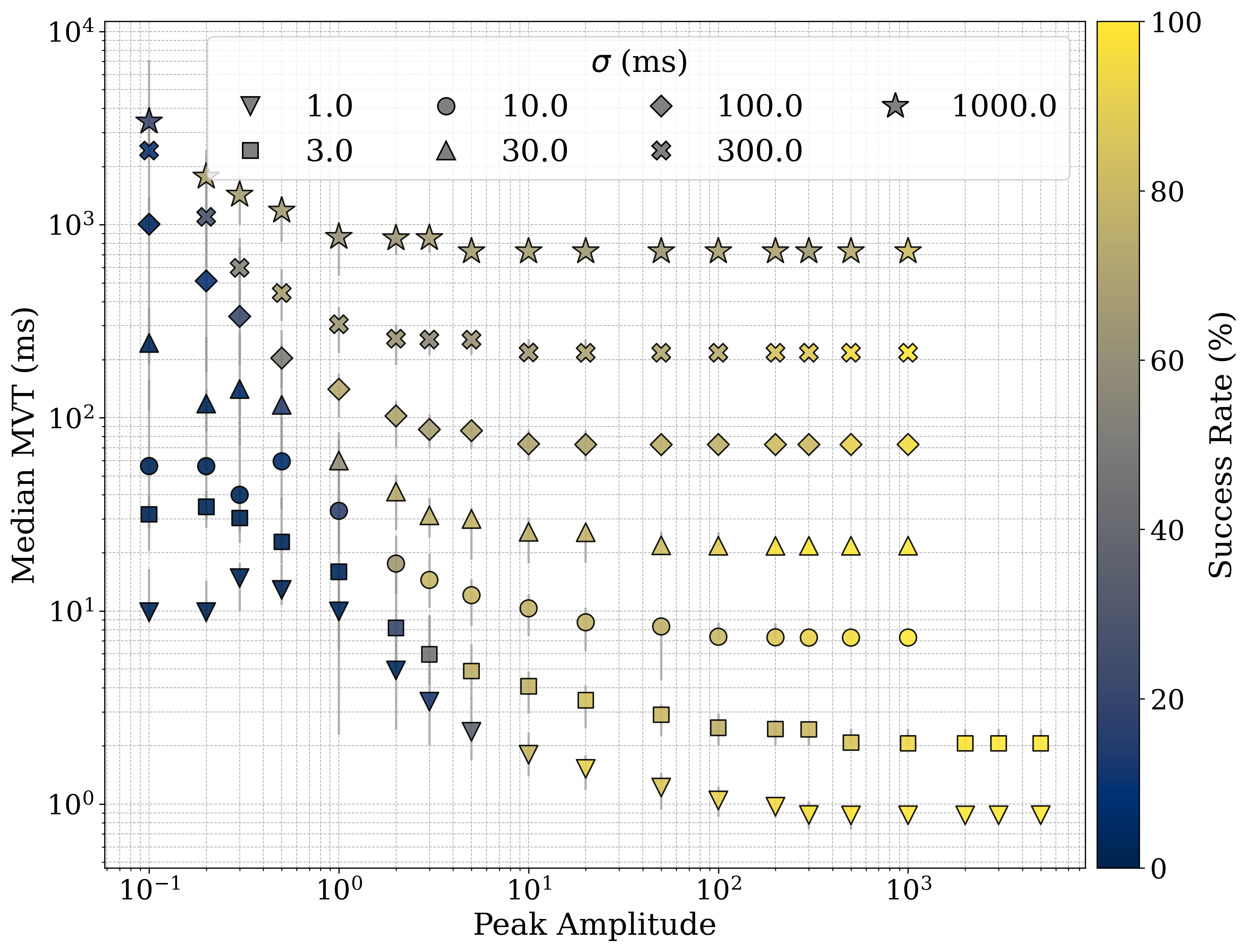}
    \caption{\small The median MVT as a function of peak amplitude, identical to Figure~\ref{fig:gaussian_mvt_vs_amp}. Here, the data points are color-coded by the measurement success rate, defined as the percentage of the 300 Monte Carlo realizations that yielded a valid MVT. This figure represents the measurement's reliability, showing a clear transition from unreliable (dark blue, low success rate) at low amplitudes to highly reliable (bright yellow, $\approx$100\% success rate) at high amplitudes.}
    \label{fig:gaussian_mvt_vs_amp_success}
\end{figure}

Figure~\ref{fig:gaussian_mvt_vs_amp} illustrates this convergence, but it does not quantify the statistical reliability of the measurements. To address this, Figure~\ref{fig:gaussian_mvt_vs_amp_success} shows the same data but color-codes each point by the success rate of the measurement. At high amplitudes (right side), the success rate is 100\% (bright yellow points), and the MVT correctly stabilizes. Conversely, at low amplitudes (left side), the success rate is near zero (dark blue points), and the few successful measurements are the scattered upper limits. Our work extends these findings by characterizing the intermediate region, showing a gradual increase in both success rate and measurement stability. Together, these figures demonstrate that a more robust metric than amplitude alone is needed to quantify a measurement's reliability.

\begin{figure}[!htpb]
    \centering
    \includegraphics[width=0.45\textwidth]{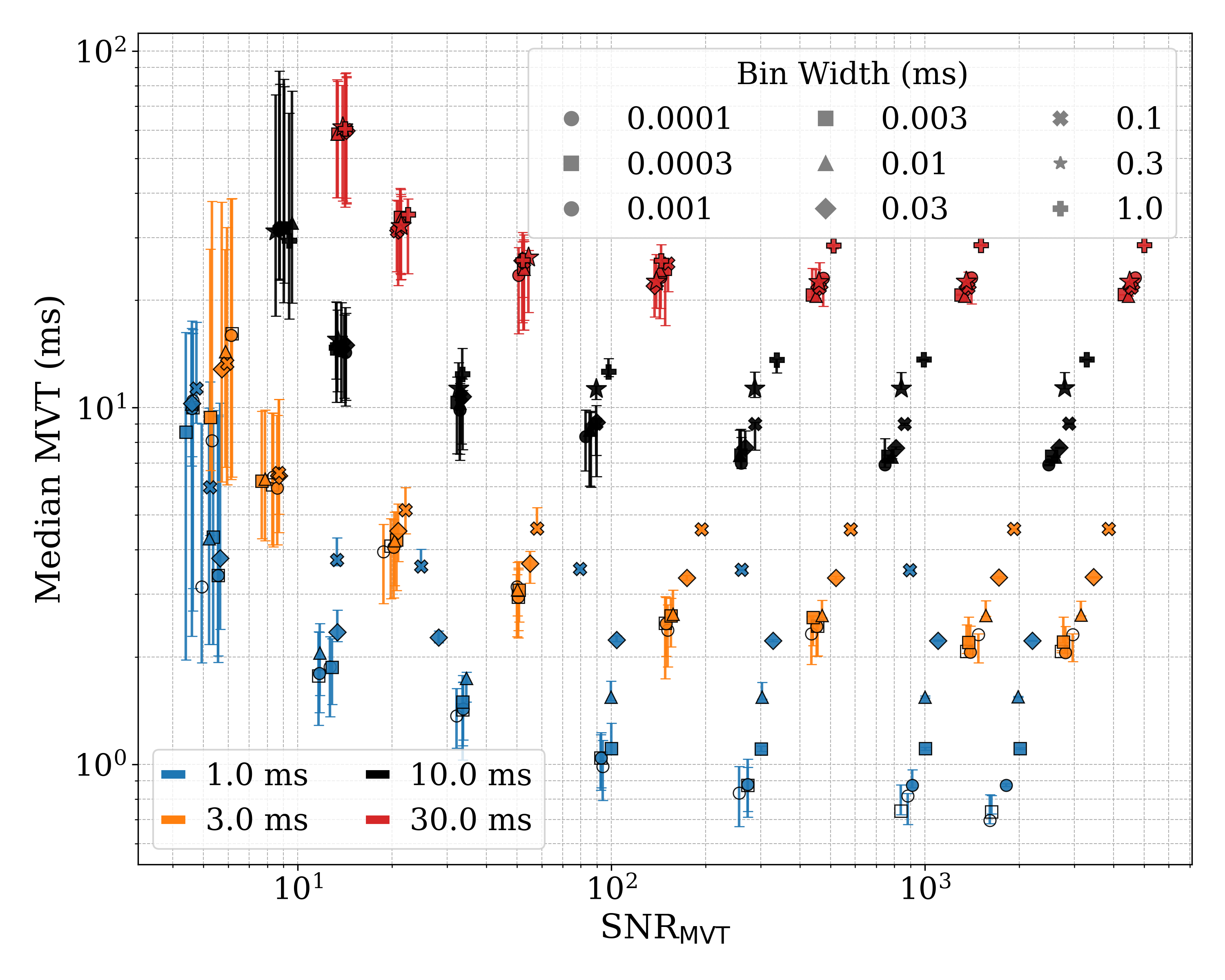}
    \caption{\small The median MVT as a function of the $\mathrm{SNR}_{\mathrm{MVT}}$ for simulated Gaussian pulses. Color indicates the intrinsic width ($\sigma$) and marker shape indicates the BW. The plot shows that MVT measurements converge from a noise-dominated, overestimated regime at low SNR to the true intrinsic timescale at high SNR. The specific SNR required for this convergence is dependent on $\sigma$, with shorter timescales requiring a higher SNR for a reliable measurement.}
    \label{fig:gaussian_mvt_vs_snr}
\end{figure}

While peak amplitude provides an intuitive proxy for signal strength, a more universal metric is needed that also accounts for the background and the timescale over which the signal is measured. 
We therefore define \textbf{$\mathrm{SNR}_{\mathrm{MVT}}$, i.e. the Signal-to-Noise Ratio computed on the timescale of the measured MVT},  as the definitive metric of a measurement's significance. Figure~\ref{fig:gaussian_mvt_vs_snr} plots the MVT against this metric for our entire suite of Gaussian simulations.

The plot shows a clear, general trend: at low $\mathrm{SNR}_{\mathrm{MVT}}$, the MVT measurements are systematically overestimated and highly scattered, whereas they converge toward the true intrinsic width, $\sigma$, at high $\mathrm{SNR}_{\mathrm{MVT}}$. However, the data show that a single, SNR threshold is an oversimplification. A closer inspection reveals that the SNR value required to achieve a reliable measurement is itself dependent on the intrinsic timescale being measured. Shorter timescales (e.g., $\sigma=1.0$ ms, blue points) require a higher SNR to converge to their true value compared to longer timescales (e.g., $\sigma=30.0$ ms, red points). 
This result establishes a key principle that we will test with more complex pulse shapes: while a general threshold of $\mathrm{SNR}_{\mathrm{MVT}} \sim 30$-$50$ serves as a useful first-order guideline, \textbf{the specific SNR required for a reliable measurement is dependent on the intrinsic timescale itself.}

\begin{figure}
    \centering
    \includegraphics[width=0.45\textwidth]{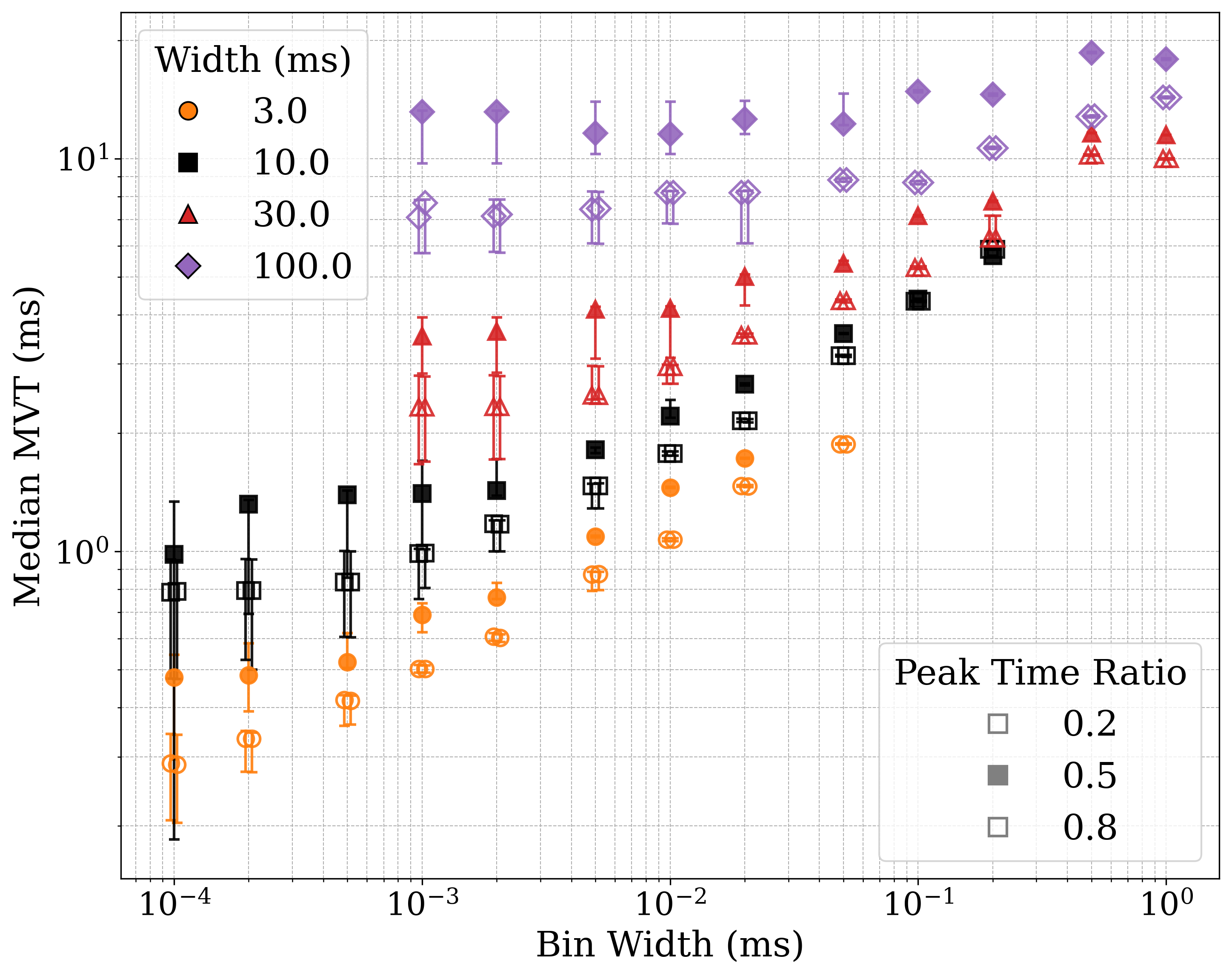}
    \caption{\small The median MVT as a function of BW for high-amplitude triangular pulses of varying total widths (colors) and rise-to-fall time ratios (marker shapes). The plot confirms the bin-limited behavior at large BWs and convergence to a stable plateau at small BWs.}
    \label{fig:triangular_mvt_vs_bw}
\end{figure}

\subsection{Robustness to Pulse Asymmetry: Triangular Pulses}
\label{subsec:triangular}

To test if our findings are robust to pulse asymmetry, we repeated the analysis for a suite of triangular pulses with varying total widths and rise-to-fall time ratios. Figure~\ref{fig:triangular_mvt_vs_bw} shows the MVT versus BW for these pulses. The plot confirms the binning effect and shows that the MVT consistently measures the timescale of the sharpest feature, which is the shorter of the rise or fall times. For example, we find a pulse with a 0.2 rise/fall ratio (fast rise) and a pulse with a 0.8 ratio (fast fall) yield similar MVT measurements, as both possess an equally short timescale component (Figure~\ref{fig:triangular_mvt_vs_bw}). This indicates that the MVT is fundamentally sensitive to the shortest timescale of variability present in the pulse profile.

\begin{figure}
    \centering
    \includegraphics[width=0.46\textwidth]{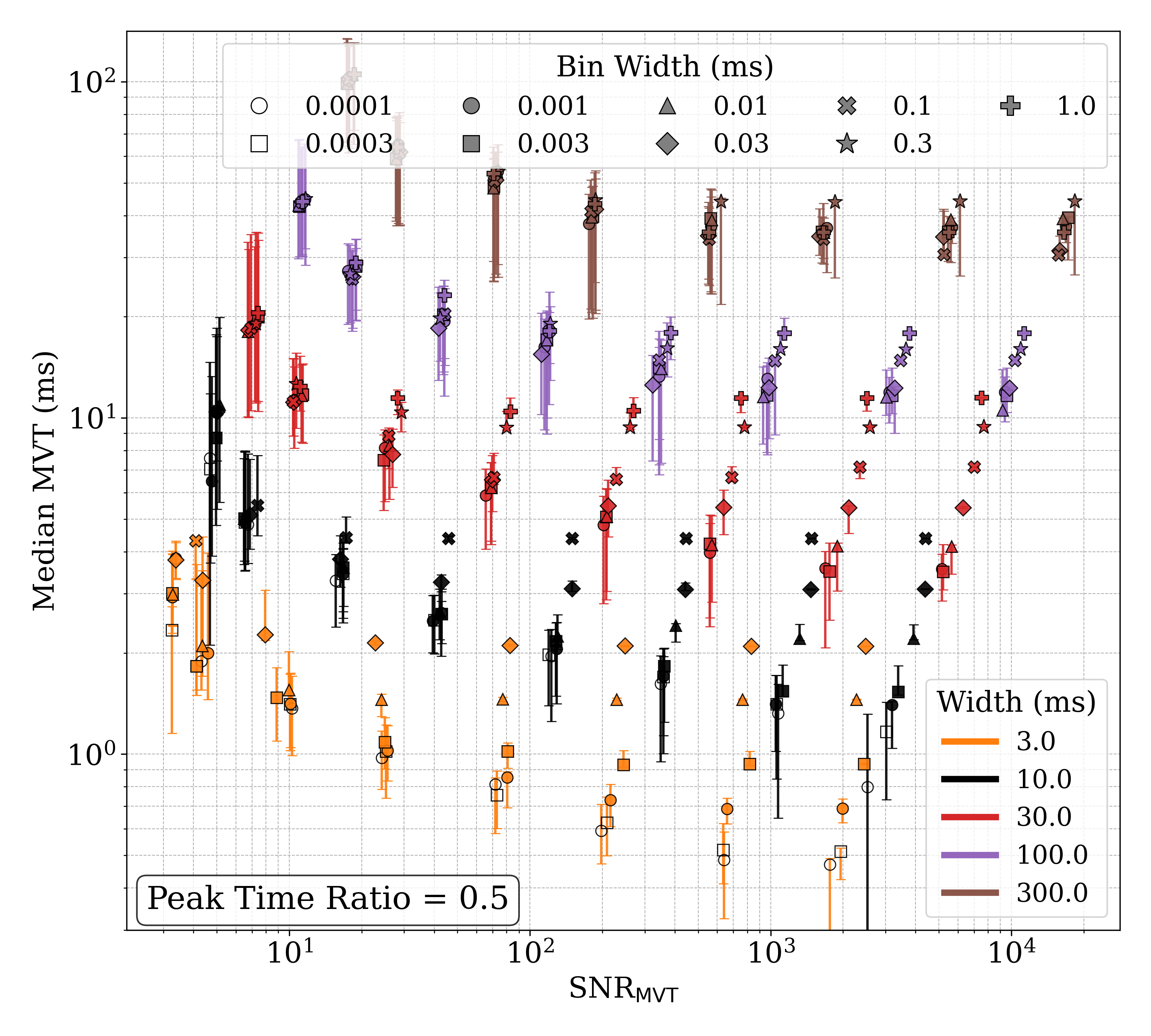}
    \caption{\small The median MVT as a function of the $\mathrm{SNR}_{\mathrm{MVT}}$ for simulated symmetric triangular pulses. Color indicates the intrinsic pulse width and marker shape indicates the analysis BW.}
    \label{fig:triangular_mvt_vs_snr}
\end{figure}

Figure~\ref{fig:triangular_mvt_vs_snr} shows the MVT as a function of $\mathrm{SNR}_{\mathrm{MVT}}$ for simulated symmetric triangular pulses. The data follow a similar track to that identified for the Gaussian pulses, with measurements being scattered and overestimated at low SNR and converging to the true intrinsic timescale at high SNR. The plot also shows the impact of the analysis BW (indicated by marker shape); measurements made with larger BWs are systematically limited and fail to converge to the shortest true timescales, even at high SNR. Consistent with the Gaussian results, we find that the SNR required for a reliable measurement is again dependent on the intrinsic pulse width, confirming that this complex relationship is a general feature and not specific to a symmetric Gaussian profile.

\subsection{Confirmation with a Realistic Model: Norris Pulses}
\label{subsec:norris}

\begin{figure}
    \centering
    \includegraphics[width=0.46\textwidth]{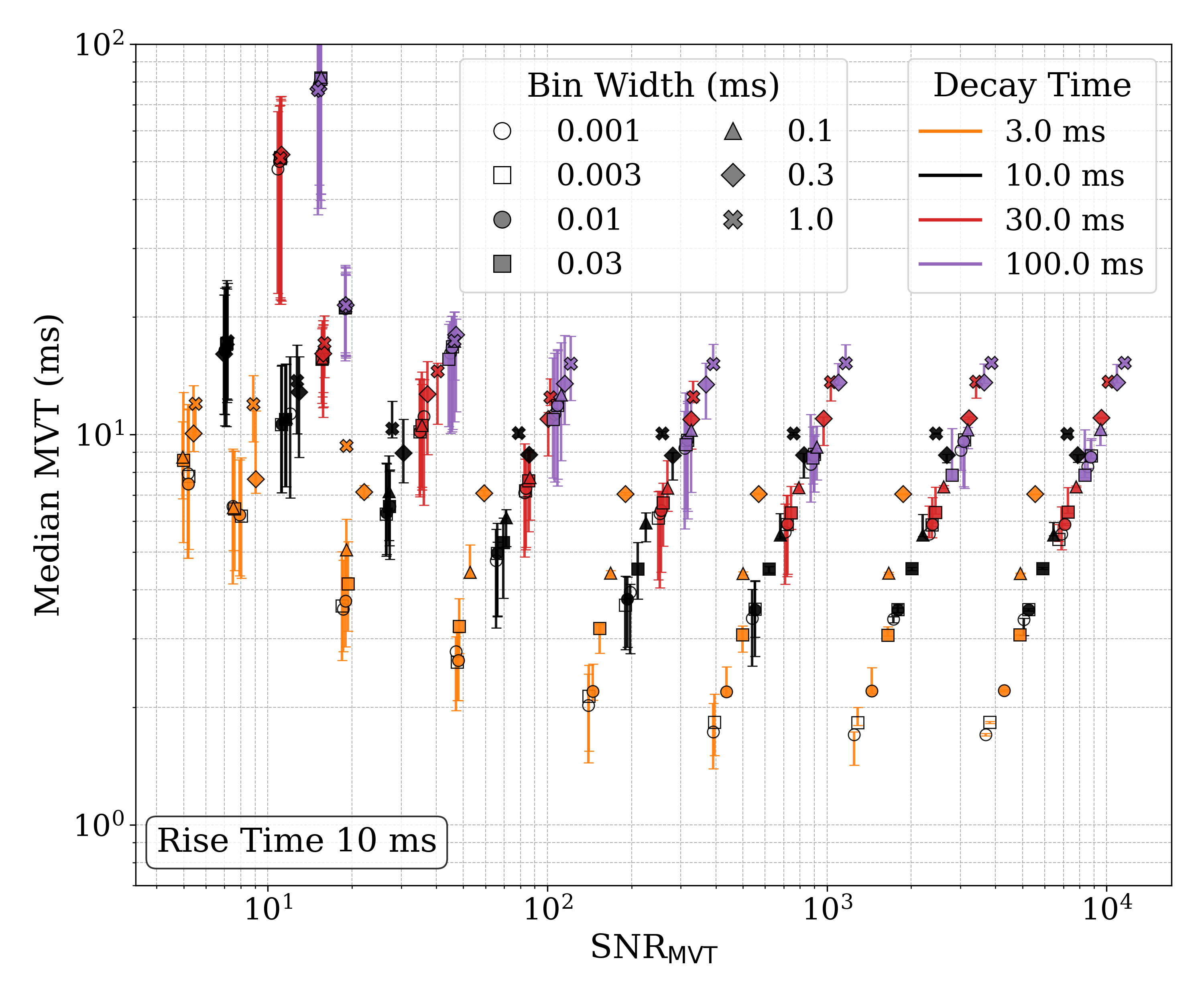}
    \caption{\small A representative example of the MVT as a function of the $\mathrm{SNR}_{\mathrm{MVT}}$ for simulated Norris pulses with a fixed rise time of 10 ms. Color indicates the pulse's decay time and marker shape indicates the analysis BW.}
    \label{fig:norris_mvt_vs_snr}
\end{figure}

As a final test, we repeated the analysis using the Norris profile, which is commonly used to model the pulses of Gamma-Ray Bursts (GRBs). Figure~\ref{fig:norris_mvt_vs_snr} shows a representative example of the results for Norris pulses with a fixed rise time of 10 ms; further examples for other rise times are provided in Appendix~\ref{appfig:norris_mvt_snr_all}. The plot confirms our findings from the simpler pulse shapes. The data once again follow the universal track, where measurements are unreliable at low SNR and converge to the true intrinsic timescale at high SNR. The plot also clearly illustrates the impact of the analysis BW (indicated by marker shape); measurements made with larger BWs are systematically limited and fail to resolve the shortest true timescales. Furthermore, \textbf{the timescale-dependent nature of the SNR threshold is preserved}, demonstrating the robustness of our framework even for physically realistic, highly asymmetric pulse profiles.

\subsection{The MVT Validation Curve}
\label{subsec:mvt_validation_curve}
\begin{figure}[htbp]
    \centering
    \includegraphics[width=0.45\textwidth]{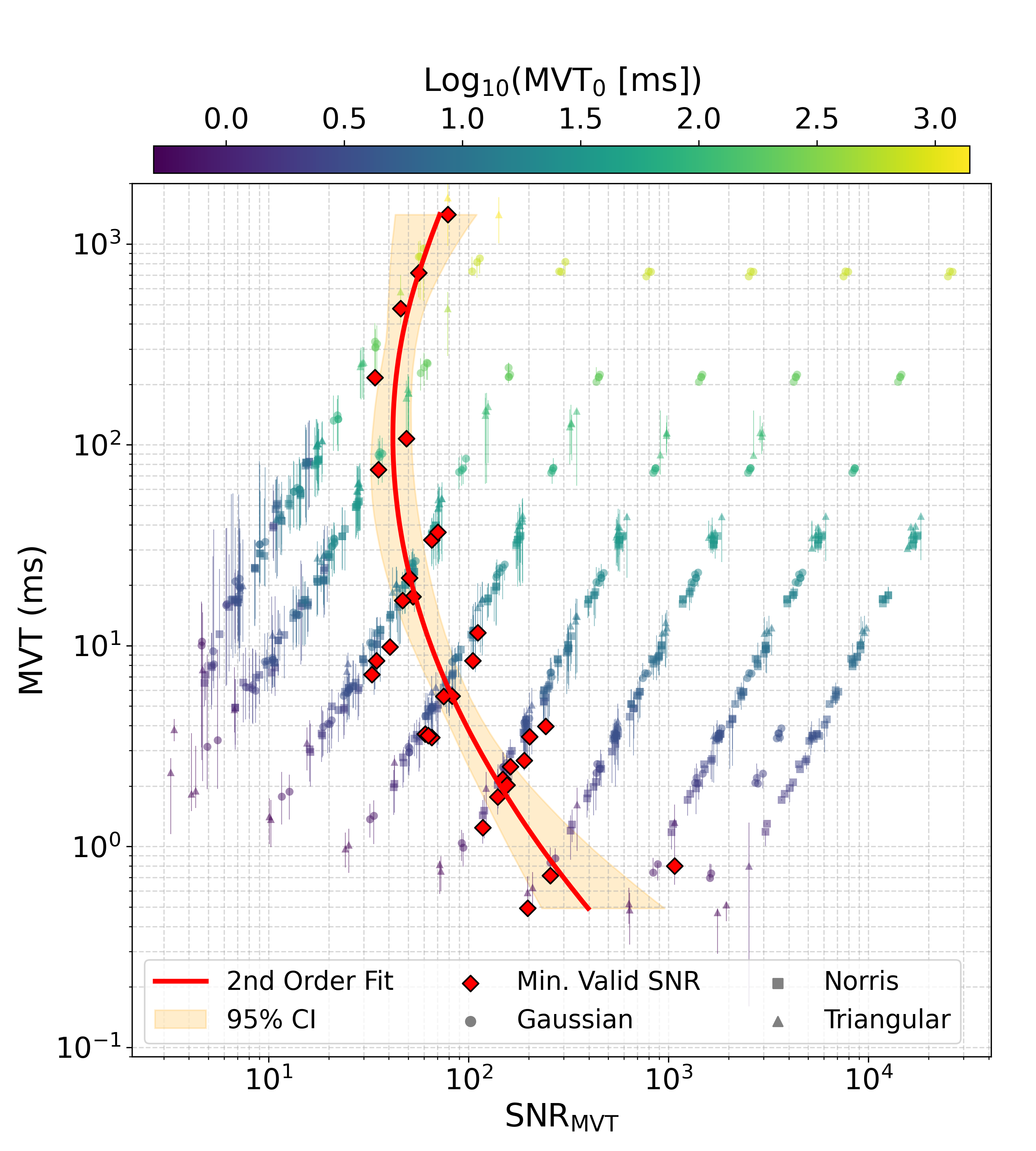}
    \caption{\small
        Measured MVT versus $\mathrm{SNR}_{\mathrm{MVT}}$ for the combined suite of isolated-pulse simulations (Gaussian, triangular, and Norris). Points are colored by their intrinsic MVT ($\mathrm{MVT}_0$). Red diamonds mark the representative “minimum-SNR” points used to define the detection boundary. The solid red curve is a second-order polynomial fit to these points, and the yellow region shows its bootstrap-derived 95\% confidence interval. This fitted curve defines the MVT Validation Curve, which sets the minimum $\mathrm{SNR}_{\mathrm{MVT}}$ required for a reliable MVT measurement.
    }
    \label{fig:mvt_validation_curve}
\end{figure}

To compare the behavior across all isolated-pulse simulations, we combine the Gaussian, triangular, and Norris results into a single diagnostic diagram (Figure~\ref{fig:mvt_validation_curve}). The plot shows the measured MVT as a function of $\mathrm{SNR}_{\mathrm{MVT}}$, with each point colored by its intrinsic timescale ($\mathrm{MVT}_0$). At low $\mathrm{SNR}_{\mathrm{MVT}}$, the measured MVT values are widely scattered and systematically biased upward. As $\mathrm{SNR}_{\mathrm{MVT}}$ increases, the measurements converge toward the true value, forming a well-defined track whose location depends on the intrinsic MVT.

The lower boundary of these converged region corresponds to the minimum $\mathrm{SNR}_{\mathrm{MVT}}$ required to recover the true MVT. We identify this boundary using a representative set of points (red diamonds). These points are fit with a second-order polynomial (solid red curve), and the uncertainty on this boundary is estimated using bootstrap resampling. The yellow shaded region shows the 95\% confidence interval on the fitted relation. We refer to this polynomial fit as the \textbf{MVT Validation Curve}. Measurements that fall to the right of this curve correspond to statistically reliable MVT determinations (R). Measurements to the left of the curve lie in the noise-dominated regime and should be reported as upper limits (UL). Points lying within the 95\% confidence band are classified as likely upper limits (LUL), since they are not statistically distinguishable from the detection threshold.

\subsection{A Systematic Workflow for MVT Interpretation}
\label{subsec:workflow}

The consistent results from our simulations of Gaussian, triangular, and Norris pulses motivate a simulation-supported workflow for interpreting MVT measurements from arbitrary light curves. This procedure evaluates both the statistical significance of a measurement and potential systematic biases introduced by temporal binning:

\begin{enumerate}
    \item \textbf{Initial Measurement:} Select an initial analysis bin width ($\mathrm{BW}_1$, e.g.\ $\sim 1$~ms) that is significantly smaller than any apparent timescales of interest in the light curve. From this light curve, measure the Minimum Variability Timescale ($\mathrm{MVT}_1$).

    \item \textbf{The Stability Check (Find the Timescale Plateau):} This step removes any systematic bias introduced by the choice of bin width.
    \begin{itemize}
        \item Create new light curves with progressively smaller bin widths ($\mathrm{BW}_2 < \mathrm{BW}_1$, $\mathrm{BW}_3 < \mathrm{BW}_2$, etc.) and re-measure the MVT ($\mathrm{MVT}_2$, $\mathrm{MVT}_3$, etc.).
        \item Continue until the measured MVT stabilizes (i.e.\ $\mathrm{MVT}_{n} \approx \mathrm{MVT}_{n-1}$ within uncertainties). This stable value, $\mathrm{MVT}_{\mathrm{final}}$, is the shortest resolvable timescale in the data.
        \item If a stable plateau cannot be achieved (the MVT continues to decrease with every reduction in bin width), the true timescale is unresolved. In this case, the 1$\sigma$ upper bound ($\mathrm{MVT}_{84}$, i.e.\ the 84th percentile of the MVT distribution at the smallest tested bin width) should be reported as an \textbf{upper limit}, and the analysis concludes here.

    \end{itemize}

    \item \textbf{The Reliability Check (Assess the Plateau):} Once the shortest stable timescale $\mathrm{MVT}_{\mathrm{final}}$ is identified, compute its signal-to-noise ratio, $\mathrm{SNR}_{\mathrm{MVT, final}}$.
    \begin{itemize}
        
        \item Compare the pair [$\mathrm{SNR}_{\mathrm{MVT, final}}, \mathrm{MVT}_{\mathrm{final}}$] to the MVT Validation Curve.
        \item If the point lies to the right of the curve, classify it as a robust measurement.
        \item If it lies below the curve, classify it as an upper limit.
        \item If it lies within the 95\% confidence region, classify it as a likely upper limit.

    \end{itemize}
\end{enumerate}

To facilitate this classification, we provide a Python tool that evaluates user-supplied ($\mathrm{MVT}, \mathrm{SNR}_{\mathrm{MVT}}$) pairs against the MVT Validation Curve, as described in the Data Availability section.

\subsection{Application to Complex, Multi-Component Profiles}
\label{subsec:complex}

\begin{figure}[ht]
    \centering
    \includegraphics[width=.46\textwidth]{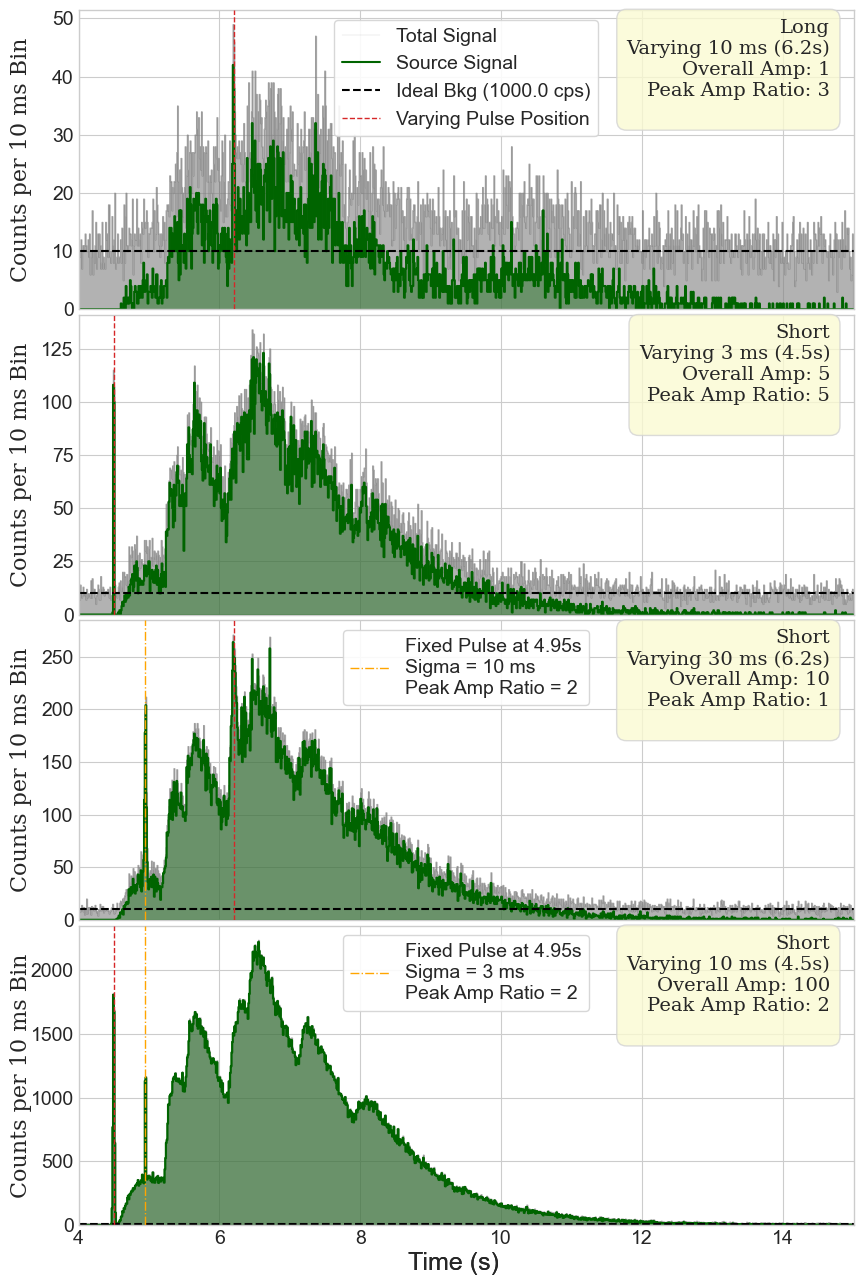} 
    \caption{\small Representative examples of the complex, multi-component light curves used for the final phase of our analysis. Each panel consists of an underlying pulse template (green), a constant background (dashed black line), an injected feature pulse, and the resulting total signal with Poisson noise (gray histogram). The panels illustrate the variety of scenarios tested, including varying the position and amplitude of a feature pulse on both long and short templates (top two panels) and injecting a second sharp feature to test the algorithm's ability to distinguish between two competing short timescales (bottom two panels).}
    \label{fig:complex_lc_example}
\end{figure}

Having established the MVT's behavior for isolated pulses, we performed a final test to simulate a more realistic astrophysical scenario where multiple structures overlap. We created two underlying templates, a \textit{Complex-Long} and a \textit{Complex-Short} lightcurve (see Section~\ref{subsec:pulses}), upon which we superimposed a much sharper Gaussian "feature". Figure~\ref{fig:complex_lc_example} shows several representative examples of the light curves generated for this analysis.
\par

\begin{figure*}[!htbp]
    \centering
    \rotatebox{90}{%
        \begin{minipage}{0.99\textheight}  
            \centering
            \includegraphics[width=0.99\textheight]{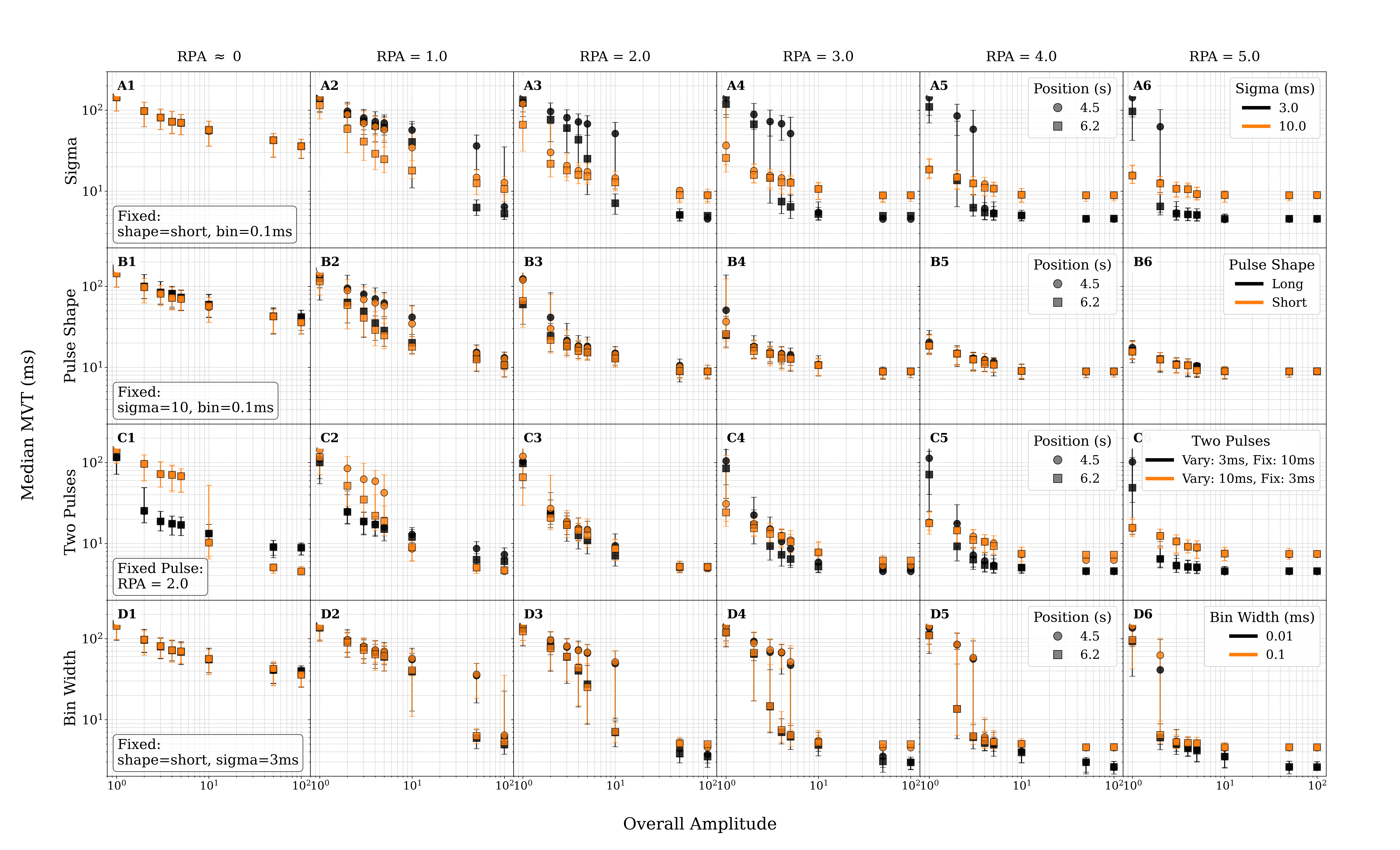}
            \
    \caption{\footnotesize{A comprehensive overview of MVT parameter dependencies. Each panel plots Median MVT vs. Overall Amplitude. The columns represent the increasing relative strength of an injected feature pulse, from  $\sim$~0\% (left) to 500\% (right). 
    \textbf{(Row A)} The dependence on the intrinsic pulse width ($\sigma$) and position, showing that wider pulses ($\sigma=10$ ms, orange squares) consistently yield longer MVTs than narrower pulses ($\sigma=3$ ms, black circles). Furthermore, for a given $\sigma$, the measured MVT is found to be independent of the pulse's position.
    \textbf{(Row B)} A comparison between the MVT measured from two different underlying templates, a `Complex-Long` (orange) and a `Complex-Short` (black). The measured MVT is found to be independent of the overall emission structure.
    \textbf{(Row C)} The algorithm's ability to distinguish between two different injected feature pulses, demonstrating its fidelity in measuring the sharpest component of the signal. The measured MVT is found to be dominated by the pulse with more power or amplitude.
    \textbf{(Row D)} The critical impact of the analysis BW. At high amplitudes, a larger BW (0.1 ms, orange squares) systematically overestimates the MVT, while a smaller BW (0.01 ms, black circles) successfully resolves the shorter timescale.
    Across all scenarios, the MVT converges to stable values as the signal strength increases.}}
            \label{fig:complex_mvt_vs_all_parameters}
        \end{minipage}
    }
\end{figure*}

Figure~\ref{fig:complex_mvt_vs_all_parameters} summarizes the complete results of this suite of simulations. The figure shows the measured MVT as a function of the overall signal amplitude (x–axis) and the relative peak amplitude (RPA) of the injected feature pulse (columns). Panel~A shows the dependence on the intrinsic width of the feature ($\sigma$): wider pulses ($\sigma = 10$~ms; orange squares) yield systematically larger MVTs than narrower pulses ($\sigma = 3$~ms). Panel~B compares two different underlying templates (`Complex–Long`, orange; `Complex–Short`, black), showing that the MVT is largely insensitive to the global envelope of the emission. Panel~C demonstrates the algorithm's ability to distinguish between two distinct injected feature pulses. Panel~D highlights the impact of the analysis bin width: larger BWs systematically overestimate the MVT at high amplitudes.

 At low overall amplitudes, or when the feature's relative amplitude is small, the sharp pulse is not statistically significant enough to be distinguished from the fluctuations in the bright, underlying pulse. In this regime, the MVT algorithm correctly identifies the shortest significant timescale of the broad template structure. It is important to note that for the \textit{Complex-Long} template this measured timescale ($\sim$140 ms) is significantly longer than that of its fastest individual component ($\sim$50 ms), due to the smoothing effect of overlapping multiple pulses.

As the feature's amplitude or the overall amplitude increases, a transition occurs. The MVT begins to detect the sharper variations, and the measured timescale decreases from that of the broad template, often through an intermediate regime where the MVT is a composite value influenced by both components. Only when the feature's relative amplitude is large does the measured MVT finally converge to the true, short intrinsic timescale of the feature pulse itself. This behavior is observed even when the feature pulse is placed at a position (e.g., at t=4.5 s) where it does not physically overlap with the main bright emission, demonstrating that the MVT algorithm assesses the statistical significance of variability across the entire observation window.

To connect these findings directly to our established framework, we plot the same data as a function of the $\mathrm{SNR}_{\mathrm{MVT}}$ in Appendix Figure~\ref{appfig:complex_mvt_vs_all_parameters_app}. It is important to clarify how this metric is calculated: the $\mathrm{SNR}_{\mathrm{MVT}}$ here represents the significance of the entire combined signal (template + feature) over the full analysis window. 
This global metric may not always reflect the true, local SNR of the specific feature pulse responsible for the MVT. However, Figure~\ref{appfig:complex_mvt_vs_all_parameters_app} demonstrates that this global $\mathrm{SNR}_{\mathrm{MVT}}$ still serves as an excellent proxy for the overall signal significance. 
The general trend holds true: the transition from measuring the template to resolving the feature consistently occurs as the overall signal significance increases, reinforcing the fundamental link between statistical significance and MVT behavior.
A full, time-resolved treatment of the local SNR is beyond the scope of this paper and will be the focus of our future work.

This is the most important finding of our work: while the MVT algorithm is effective for isolated pulses, in complex signals its interpretation requires significant caution.
We find that a measured MVT can be an intermediate "blended" value, or it can reflect the timescale of a dominant pulse that is not necessarily the fastest one. 
This holds true even when a faint feature does not physically overlap with a brighter one, as the algorithm assesses significance across the entire observation window. This implies that many MVT measurements in the literature, particularly those from complex or low-SNR signals, should be conservatively treated as upper limits. 

\subsection{Source vs. Measured Minimum Variability}
It is essential to distinguish between the intrinsic \textit{Source MVT}, representing the physical variability of the GRB central engine, and the \textit{Measured MVT}, which is the shortest statistically significant timescale identified in the detector count data. While changing the source spectrum or detector orientation alters the lightcurve morphology and integrated count rates, the MVT Validation Curve remains a property of the estimator itself. 

The instrumental response—specifically photon redistribution and incidence angles—acts as a "forward-folding" process that transforms the Source MVT into the Measured MVT. As shown in our targeted simulations incorporating full Fermi-GBM responses, these effects primarily modulate signal intensity (captured by $\mathrm{SNR}_{\mathrm{MVT}}$) but do not shift the fundamental reliability boundary of the Haar-wavelet method. Consequently, we provide a robust workflow to classify the \textit{Measured MVT} as either a robust detection or an upper limit, mapping observed data to the nearest statistically significant "functional" timescale.

\section{MVT of Real GRBs Observed by Fermi-GBM}
\label{sec:real_grbs}

To demonstrate the application of our method, we analyze a small sample of Fermi-GBM GRBs. These bursts are chosen covering a range in brightness, inferred MVT, and classification context.
The analysis is performed on publicly available Fermi Gamma-ray Burst Monitor (GBM) data using the \hyperlink{https://astro-gdt.readthedocs.io/projects/astro-gdt-fermi/en/latest/index.html}{GDT-Fermi}\footnote{\url{https://astro-gdt.readthedocs.io/projects/astro-gdt-fermi/en/latest/index.html}} \citep{GDT-Fermi}. For each GRB, we define a source interval and two background intervals (one before and one after the burst), with the specific time selections for our sample detailed in Table~\ref{apptab:fermi_gbm_details}.

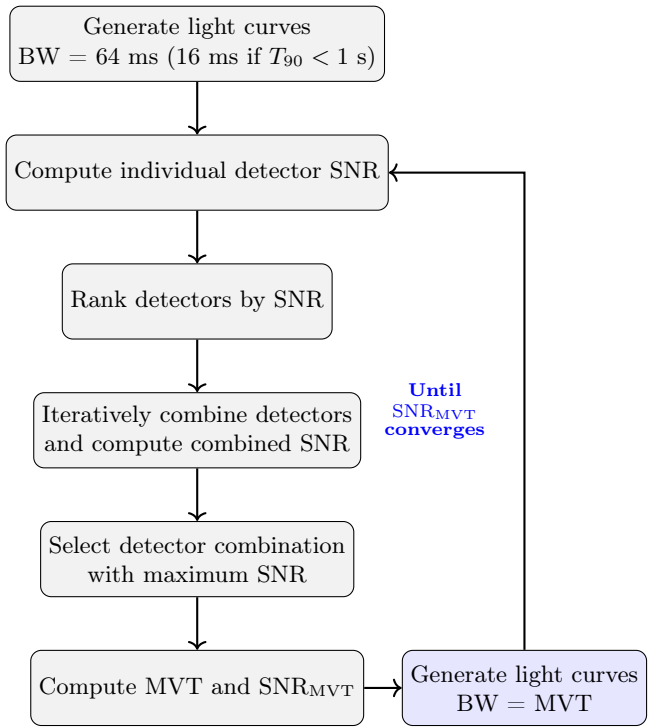
\begin{figure}[t]
\centering
\begin{tikzpicture}[
    node distance=0.7cm,
    process/.style={
        rectangle, rounded corners,
        draw=black,
        fill=black!5,
        align=center,
        minimum width=3.cm,
        minimum height=1cm
    },
    arrow/.style={->, thick}
]

\node (lc) [process] {Generate light curves\\
BW = 64 ms (16 ms if $T_{90}<1$ s)};
\node (snr) [process, below=of lc] {Compute individual detector SNR};
\node (rank) [process, below=of snr] {Rank detectors by SNR};
\node (combine) [process, below=of rank] {Iteratively combine detectors\\
and compute combined SNR};
\node (select) [process, below=of combine] {Select detector combination\\
with maximum SNR};
\node (check) [process, below=of select] {Compute MVT and  $\mathrm{SNR}_{\mathrm{MVT}}$};
\node (update) [process, right=0.5cm of check, fill=blue!10] {Generate light curves\\
BW = MVT};

\draw [arrow] (lc) -- (snr);
\draw [arrow] (snr) -- (rank);
\draw [arrow] (rank) -- (combine);
\draw [arrow] (combine) -- (select);
\draw [arrow] (select) -- (check);

\draw [arrow] (check.east) -- (update.west);
\draw [arrow] (update.north) |- 
node[pos=0.25,left=4mm,font=\scriptsize\bfseries, color = blue] 
{\shortstack{Until \\ $\mathrm{SNR}_{\mathrm{MVT}}$ \\ converges}} 
(snr.east);

\end{tikzpicture}
\caption{Flowchart illustrating the iterative detector selection procedure used to maximize the signal-to-noise ratio for each GRB. The loop indicates that light curves are regenerated using the measured MVT until the SNR converges.}
\label{fig:detector_selection_flowchart}
\end{figure}

\begin{figure}[!htpb]
    \centering
    \includegraphics[width=0.47\textwidth]{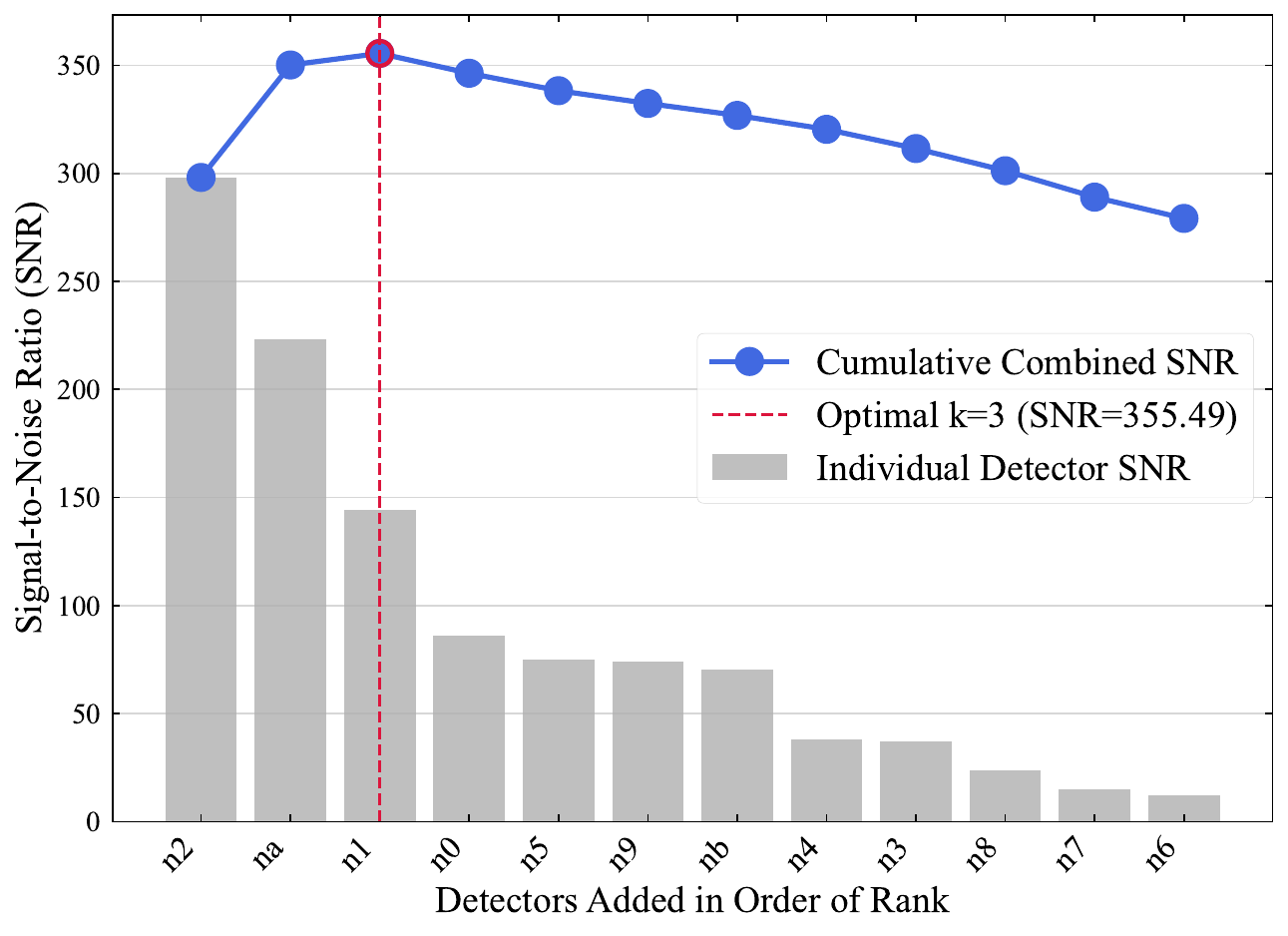} 
    \caption{An example of the iterative detector selection process for GRB 211211A (trigger bn211211549). Detectors are ranked by their individual SNR and added cumulatively one by one. The total combined SNR peaks at k=3 detectors, which are then selected for the final analysis. Adding further, lower-SNR detectors degrades the overall signal quality.}
    \label{fig:snr_vs_ndet}
\end{figure}

To select the optimal combination of detectors, we employ an iterative signal-to-noise ratio (SNR) maximization procedure. First, light curves are generated for each available NaI detector, using a BW of 64~ms (or 16~ms for bursts with $T_{90}<1$~s). The detectors are then ranked from highest to lowest individual SNR. Next, we create a series of combined light curves, iteratively adding one detector at a time in order of its rank, and calculate the new combined SNR at each step. An example of this procedure is shown in Figure~\ref{fig:snr_vs_ndet}. The optimal detector combination is the one that yields the maximum combined SNR.

For each GRB, we generated a final light curve from the combined TTE data of its optimal detector combination, typically using a BW of 0.1~ms. To estimate the MVT and its uncertainty, we treat the observed counts in each bin as the mean of a Poisson distribution and generate 300 independent Monte Carlo realizations. The MVT is calculated for each of these 300 statistical samples using the GB14 method, creating a distribution of MVT values. We then repeat the process of finding optimal detector combination with the measured MVT (Figure~\ref{fig:detector_selection_flowchart}). If we find a significant change in the $\mathrm{SNR}_{\mathrm{MVT}}$, we use the new combination. The final measurement is reported as the median of this distribution, with the $1\sigma$ uncertainties derived from the 16th and 84th percentiles. This iterative procedure ensures the measurements are internally consistent and reproducible.

\begin{table}[ht]
    \centering
    \setlength{\tabcolsep}{1.5pt} 
    \caption{MVT Measurement Results for the Fermi-GBM Sample. The assessment is determined by applying the workflow developed in this paper and is abbreviated as follows: R (Robust Measurement), LUL (Likely Upper Limit), UL (Upper Limit), and Bin-Lim. (Bin-Limited). Full details for each analysis are in the Appendix \ref{apptab:fermi_gbm_details}.}
    \label{tab:fermi_gbm_examples}
    \begin{tabular}{lcccc}
    \toprule
    GRB & Bin Width & Median MVT & $\mathrm{SNR}_{\mathrm{MVT}}$ & Assessment \\
     & BW (ms) & (ms) & & \\
    \midrule
    \textbf{211211A} & 0.1 & $4.7_{-0.7}^{+0.9}$ & 121.9 & Bin-Lim. \\
     & 0.01 & $3.6_{-0.8}^{+1.6}$ & 113.5 & Stable \\
     & 0.003 & $3.9_{-1.0}^{+1.0}$ & 116.4 & \textbf{LUL} \\
    \midrule
    \textbf{230307A} & 0.1 & $4.4_{-0.6}^{+0.7}$ & 139.7 & Bin-Lim. \\
     & 0.01 & $1.9_{-0.4}^{+0.7}$ & 94.2 & Stable \\
     & 0.003 & $1.7_{-0.4}^{+0.7}$ & 90.9 & \textbf{UL} \\
    \midrule
    \textbf{170817A} & 1.0 & $361.6_{-64.3}^{+85.9}$ & 7.45 & Stable  \\
     & 0.1 & $344.6_{-141.7}^{+133.9}$ & 7.17 & \textbf{UL}\\
    \midrule
    \textbf{231115A} & 0.1 & $10.5_{-1.8}^{+2}$ & 24.19 & \textbf{UL} \\
     & 0.01 & $9_{-1.9}^{+2.9}$ & 24.11 & UL \\
    \midrule
    \textbf{250919A} & 0.1 & $29.3_{-4.2}^{+7.4}$ & 184.41 & Stable \\
     & 0.01 & $28.8_{-5.5}^{+10.7}$ & 180.51 & \textbf{R} \\
    \bottomrule
    \end{tabular}
    \label{tab:Fermi_GRB_Short_table}
\end{table}

\begin{figure}[htbp]
    \centering
    \includegraphics[width=0.45\textwidth]{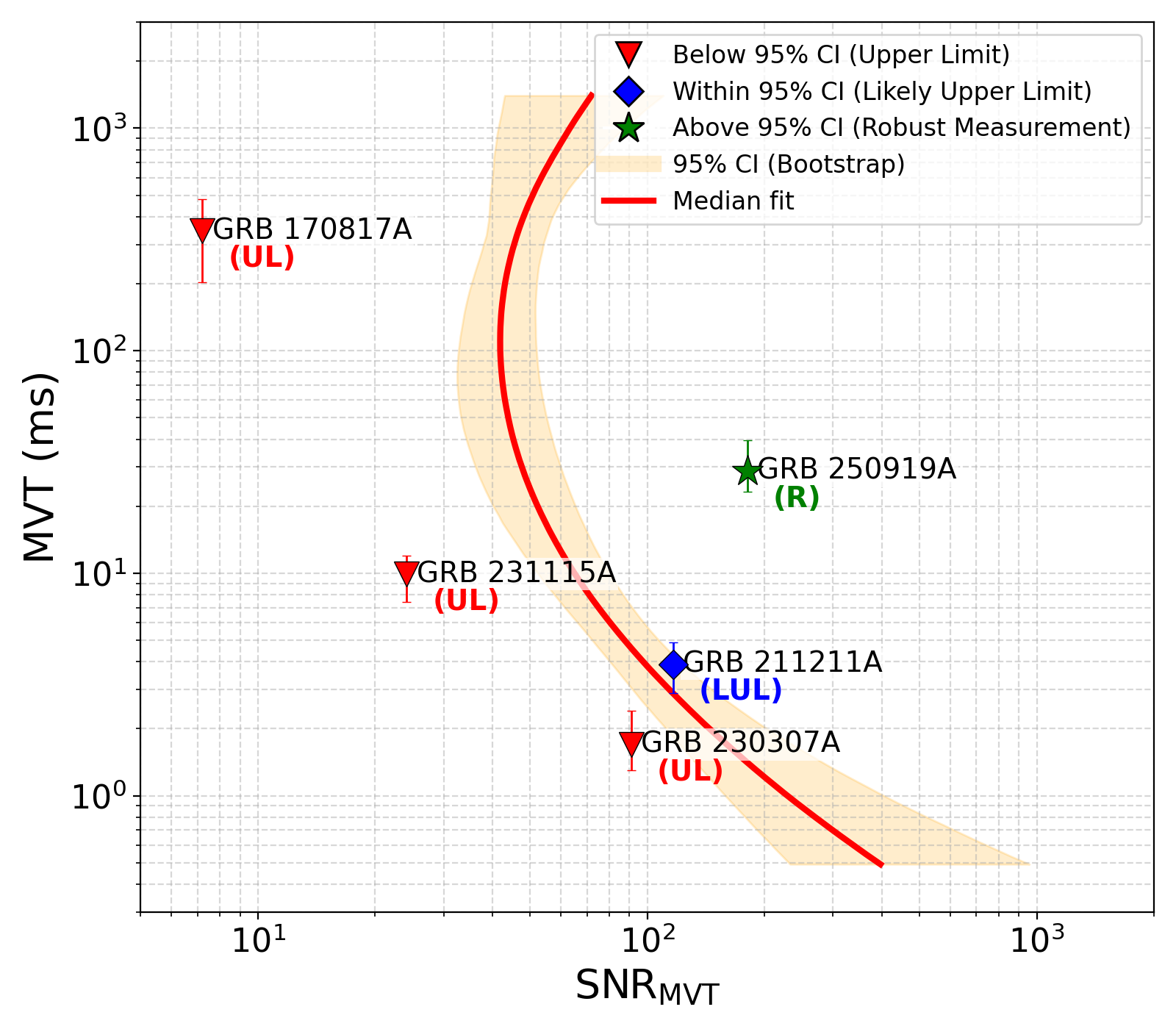}
    \caption{\small
        Application of the MVT Validation Curve to the \textit{Fermi}-GBM GRB sample. Each point shows the measured MVT and $\mathrm{SNR}_{\mathrm{MVT}}$ for a burst (with error bars), plotted against the median Validation Curve (solid red line) and its 95\% confidence interval (shaded orange region), as derived from our simulation suite. The color of each point indicates its classification based on its position relative to the Validation Curve: robust measurement (green; GRB 250919A), likely upper limit (blue; GRB 211211A), and upper limit (red; GRB 170817A, GRB 231115A, GRB 230307A). This figure provides the visual basis for the classifications listed in Table~\ref{tab:Fermi_GRB_Short_table}.
    }
    \label{fig:GRB_on_Min_SNR_Curve}
\end{figure}

Following the iterative procedure described above, we now detail the MVT results for our sample. Before presenting the results for individual GRBs, it is essential to address a conceptual limitation regarding our significance metric. As demonstrated in our complex light-curve simulations (Section \ref{subsec:complex}), it is possible for a light curve to exhibit a high overall $\mathrm{SNR}_{\mathrm{MVT}}$ driven by a bright, slowly varying envelope, while a faint but rapid feature elsewhere may possess a lower local SNR and remain undetected. As shown in Fig. \ref{appfig:complex_mvt_vs_all_parameters_app}, in the high-$\mathrm{SNR}_{\mathrm{MVT}}$ regime, the algorithm generally follows the MVT corresponding to the fastest statistically significant feature. However, because the measured $\mathrm{SNR}_{\mathrm{MVT}}$ is a global metric calculated over the entire analysis interval, the actual SNR corresponding to the specific MVT feature is most probably lower than the measured global value. While a full time-resolved treatment of local SNR is beyond the scope of this work and remains a focus for future development, this distinction is vital for interpreting the following results. It implies that the recovered MVT represents the most statistically dominant scale of variability within the analyzed window, rather than necessarily the absolute fastest intrinsic feature, and the classification based on $\mathrm{SNR}_{\mathrm{MVT}}$ represents the best possible (upper bound) significance value.

\textbf{GRB 211211A:} We first apply our workflow to GRB~211211A, a bright, long–duration burst. Our initial analysis used a 0.1~ms binned light curve constructed from the optimal combination of three detectors (N2, NA, N1), yielding a median MVT of $4.7_{-0.7}^{+0.9}$~$\mathrm{ms}$. For comparison, using only the single best detector (N2) at the same bin width resulted in a substantially longer MVT of $8.8_{-1.7}^{+2.0}$~$\mathrm{ms}$, illustrating the importance of combining detectors to maximize the statistical significance.

Because the initial SNR was high, we proceeded to the Stability Check by reducing the bin width. At 0.01~ms, we measure $3.6_{-0.8}^{+1.6}$~$\mathrm{ms}$, and at 0.003~ms we measure $3.9_{-1.0}^{+1.0}$~$\mathrm{ms}$ with $\mathrm{SNR}_{\mathrm{MVT}} \approx 116$. The decrease in MVT as the bin width is reduced confirms that the original 0.1~ms value was bin–limited, and that the shortest stable value lies at $\sim$3.6–3.9~ms.

Finally, this final MVT lies \emph{within} the 95\% confidence region of the MVT Validation Curve. Therefore, following our classification scheme, this measurement is categorized as a \textbf{likely upper limit.}

\textbf{GRB 230307A:} Our initial analysis of GRB~230307A used the optimal five–detector combination at 0.1~ms, yielding a median MVT of $4.4_{-0.6}^{+0.7}$~$\mathrm{ms}$ with $\mathrm{SNR}_{\mathrm{MVT}} \approx 137$. Because this initial SNR was high, we carried out the Stability Check by reducing the bin width to 0.01~ms and 0.003~ms. At 0.01~ms the measured MVT decreases to $1.9_{-0.4}^{+0.7}$~$\mathrm{ms}$, and at 0.003~ms it remains consistent at $1.7_{-0.4}^{+0.7}$~$\mathrm{ms}$ with $\mathrm{SNR}_{\mathrm{MVT}} \approx 91$. This confirms that the initial 0.1~ms value was bin–limited, and that the shortest stable timescale lies near $\sim$1.7–1.9~ms.

Applying the Reliability Check, this final measurement lies to the left of the MVT Validation Curve. Therefore, following our classification scheme, we report the $1\sigma$ upper bound ($\mathrm{MVT}_{84} \simeq 2.4$~$\mathrm{ms}$) as the \textbf{upper limit.}

\textbf{GRB 231115A:} GRB~231115A is an intense short burst, possibly associated with a magnetar giant flare in M82 \citep{2025A&A...694A.323T}. Using the optimal combination of six detectors, our initial 0.1~ms analysis yields a median MVT of $10.5_{-1.8}^{+2.0}$~$\mathrm{ms}$ with $\mathrm{SNR}_{\mathrm{MVT}} \approx 24$. Reducing the bin width to 0.01~ms produces a comparable value of $9.0_{-1.9}^{+2.9}$~$\mathrm{ms}$, indicating that the measurement is not strongly bin–limited. However, in the Reliability Check, both of these measurements fall to the left of the MVT Validation Curve, indicating insufficient $\mathrm{SNR}_{\mathrm{MVT}}$ for this timescale regime. Therefore, following our classification scheme, we report the $1\sigma$ upper bound at 0.01~ms ($\mathrm{MVT}_{84} \simeq 12$~$\mathrm{ms}$) as the \textbf{upper limit} on the intrinsic variability.

\textbf{GRB 170817A:} GRB~170817A is the electromagnetic counterpart to the first binary–neutron–star merger detected in gravitational waves \citep{LIGO-Fermi17}. Using the optimal combination of six detectors, our initial 1.0~ms analysis yields a median MVT of $361.6_{-64.3}^{+85.9}$~$\mathrm{ms}$. Reducing the bin width to 0.1~ms gives $344.6_{-141.7}^{+133.9}$~$\mathrm{ms}$, which is statistically consistent with the 1.0~ms value, indicating that the measurement is not strongly bin–limited. However, both measurements have extremely low signal–to–noise ratios ($\mathrm{SNR}_{\mathrm{MVT}} \approx 7.5$ and $7.2$), placing them well below the reliability threshold defined by the MVT Validation Curve. Therefore, following our classification scheme, we report the $1\sigma$ upper bound ($\mathrm{MVT}_{84} \simeq 450$~$\mathrm{ms}$) as the \textbf{upper limit.}

\textbf{GRB 250919A:} GRB~250919A provides an example of a well–resolved MVT measurement. Using the optimal combination of four detectors at 0.1~ms, our initial analysis yields a median MVT of $29.3_{-4.2}^{+7.4}$~$\mathrm{ms}$ with $\mathrm{SNR}_{\mathrm{MVT}} \approx 184$. To perform the Stability Check, we reduced the bin width to 0.01~ms, which yields $28.8_{-5.5}^{+10.7}$~$\mathrm{ms}$. This value is statistically consistent with the 0.1~ms result, confirming that a stable plateau has been reached and that the measurement is not bin–width limited.

Notably, the standard single–realization MVT calculation (“Single MVT”) did not return a significant value for this configuration, further demonstrating the importance of Monte Carlo realizations for obtaining reliable estimates. Finally, in the Reliability Check, the stable value lies well to the right of the MVT Validation Curve. Therefore, we classify the $\sim$29~$\mathrm{ms}$ timescale as a \textbf{robust measurement.}

The results for these key examples, along with the analysis of our full GRB sample, are summarized in Table~\ref{tab:fermi_gbm_examples}

\section{Discussion and Conclusion}
\label{sec:discussion}

In this work, we build on GB14 by using a comprehensive suite of simulations to establish a quantitative framework for validating Haar--based MVT measurements. Across a diverse set of isolated pulse shapes, the MVT is not tied to any single pulse parameter, but instead acts as a pulse--shape–independent proxy for the shortest statistically significant timescale present in the emission.

To translate this finding into a practical diagnostic tool, we aggregated our simulation results (Gaussian, triangular, and Norris) into a single MVT–$\mathrm{SNR}_{\mathrm{MVT}}$ diagnostic plot (Figure~8). By fitting the lower boundary, of the converged, high–significance measurements, we defined the \textbf{MVT Validation Curve}. This curve provides a quantitative threshold specifying the minimum $\mathrm{SNR}_{\mathrm{MVT}}$ required to reliably resolve a given timescale. We then established a formal workflow (Section~3.5) that uses this curve to classify any new measurement as a \textbf{Robust Detection, a Likely Upper Limit, or an Upper Limit} based on its position relative to this reliability boundary.

The necessity for this curve is driven by our primary finding: the reliability of an MVT measurement depends jointly on its value and on $\mathrm{SNR}_{\mathrm{MVT}}$. Faster intrinsic variability requires proportionally higher signal–to–noise to be resolved. This dependency explains the behavior we observed in multi–component light curves, where a faint rapid feature can be overshadowed by a brighter, slower one, yielding an apparently intermediate MVT even when faster variability is present. Taken together, our simulations imply two practical conclusions: (1) the Haar estimator returns the timescale of the most statistically dominant structure, which is not always the fastest intrinsic feature; and (2) in typical observational regimes, many reported MVT values may be upper limits rather than direct measurements, and therefore should be tested against the MVT Validation Curve to determine whether they satisfy the required $\mathrm{SNR}_{\mathrm{MVT}}$ threshold for a robust detection.

By establishing the MVT Validation Curve using instrument-agnostic analytic templates, we define a fundamental statistical baseline for the Haar-wavelet estimator. While instrumental smearing and energy-dependent variability transform the ``Source MVT'' into the ``Measured MVT,'' these factors primarily manifest as signal intensity modulations. Because the reconstruction of the intrinsic Source MVT from observed data constitutes an ill-posed inversion problem, analogous to the forward-folding requirement in spectral analysis, we treat these instrumental signatures as inherent properties of the Measured MVT. Thus, the MVT Validation Curve presented here serves as a robust first-generation, empirically calibrated result. We acknowledge that the precise mapping between measured and source timescales may depend on specific instrumental and spectral details; these nuances will be a primary focus of future instrument-specific catalogs and cross-instrument MVT comparative studies.

We then applied this framework to a sample of real \textit{Fermi}–GBM observations to demonstrate its practical utility. Our analysis confirmed a robust measurement in a bright event such as GRB~250919A, and enabled a re–classification of several previously ambiguous cases. For example, the short timescale reported for GRB~230307A is shown to be an upper limit due to its transitional $\mathrm{SNR}_{\mathrm{MVT}}$, while the faint signal in the landmark event GRB~170817A is classified as a noise–dominated regime and therefore reported as an upper limit.

\subsection{Interpretation MVT with Global vs. Local Significance}
It is essential to acknowledge the conceptual limitations inherent in a global variability analysis. Our $\mathrm{SNR}_{\mathrm{MVT}}$ is a global metric, calculated over the entire signal interval. Consequently, the measured MVT are often driven by a bright, slowly varying component, while a faint but rapid feature may be undetected. We have also seen from the complex, multi-component light curves, the MVT may converge to an intermediate or ``blended'' value. Most of the observed light curves are expected to behave like our complex, multi-component simulation.

It is therefore probable that in many instances, the intrinsic variability timescale is lower than the measured Global MVT. Conversely, because the global $\mathrm{SNR}_{\mathrm{MVT}}$ is calculated over the entire light curve, the actual SNR corresponding to a localized MVT feature is likely lower than the measured global value. Together, this indicates that the current classifications presented in this work represent the best-case (upper bound) significance values for these detections. This motivates the next logical step: a time-resolved MVT analysis. Evaluating the MVT in sliding temporal windows will likely recover shorter timescales and lower local $\mathrm{SNR}_{\mathrm{MVT}}$ values, which could \textbf{refine physical constraints by potentially reducing the inferred emission region sizes, $R \approx 2\,c\,\Gamma^2\,t_{\mathrm{MVT}}$, even further.}

In conclusion, the framework presented here provides a foundation for standardizing MVT analysis and for placing individual measurements in a physically interpretable context. By demonstrating the caution needed when interpreting variability in complex signals, we have provided both a practical workflow for immediate use and a clear path forward for future studies of the rapid variability that traces the central engines of the most energetic events in the Universe.

\section*{Acknowledgments}
The USRA coauthors gratefully acknowledge NASA funding from cooperative agreement 80NSSC24M0035.
The UAH coauthors gratefully acknowledge NASA funding from cooperative agreement 80MSFC22M0004.
The UAH coauthors gratefully acknowledge the Alabama Supercomputer Authority for providing computational resources and support that contributed to the results reported in this paper.
The coauthors from IISER Thiruvananthapuram acknowledges the High Performance Computing Center (CHPC) of IISER Thiruvananthapuram for providing computational resources. We gratefully acknowledge the \textit{Fermi} GBM team for maintaining and operating the instrument, and the public availability of GBM data products. We thank colleagues and collaborators, especially Dr.\ Narayan Bhat and Prof.\ Nathaniel Butler, for discussions and feedback during the development of this work. We also acknowledge the use of language–editing tools during manuscript preparation. We would like to thank the anonymous reviewer for their insightful feedback and for help in improving the manuscript.

\section*{Data Availability}
The \textit{Fermi}–GBM datasets used in this analysis are publicly available from the HEASARC archive\footnote{\url{https://heasarc.gsfc.nasa.gov/}}. The Gamma–ray Data Tools (\texttt{gdt}) and \texttt{gdt–fermi} software packages are available from their documentation site\footnote{\url{https://astro-gdt.readthedocs.io/en/latest/index.html}}. The simulation code used to generate the results in this work is available on GitHub\footnote{\url{https://github.com/sumanbala2210-USRA/GBM_MVT_paper}}. This repository includes the final MVT–$\mathrm{SNR}_{\mathrm{MVT}}$ fit model (\texttt{mvt\_snr\_fit\_model.npz}) and a Python helper script (\texttt{classify\_mvt\_value.py}) that allows users to test their own MVT measurements against the validation curve.

\bibliography{GRB_complete}
\bibliographystyle{aasjournalv7}

\appendix
\section{Supplementary Plots and Data}

This appendix provides supplementary materials that support the analyses and findings presented in the main paper.

\subsection{Gaussian Pulse Simulations}
\label{appsec:gaussian_sims}

\begin{figure*}[ht]
    \centering
    \includegraphics[width=\textwidth]{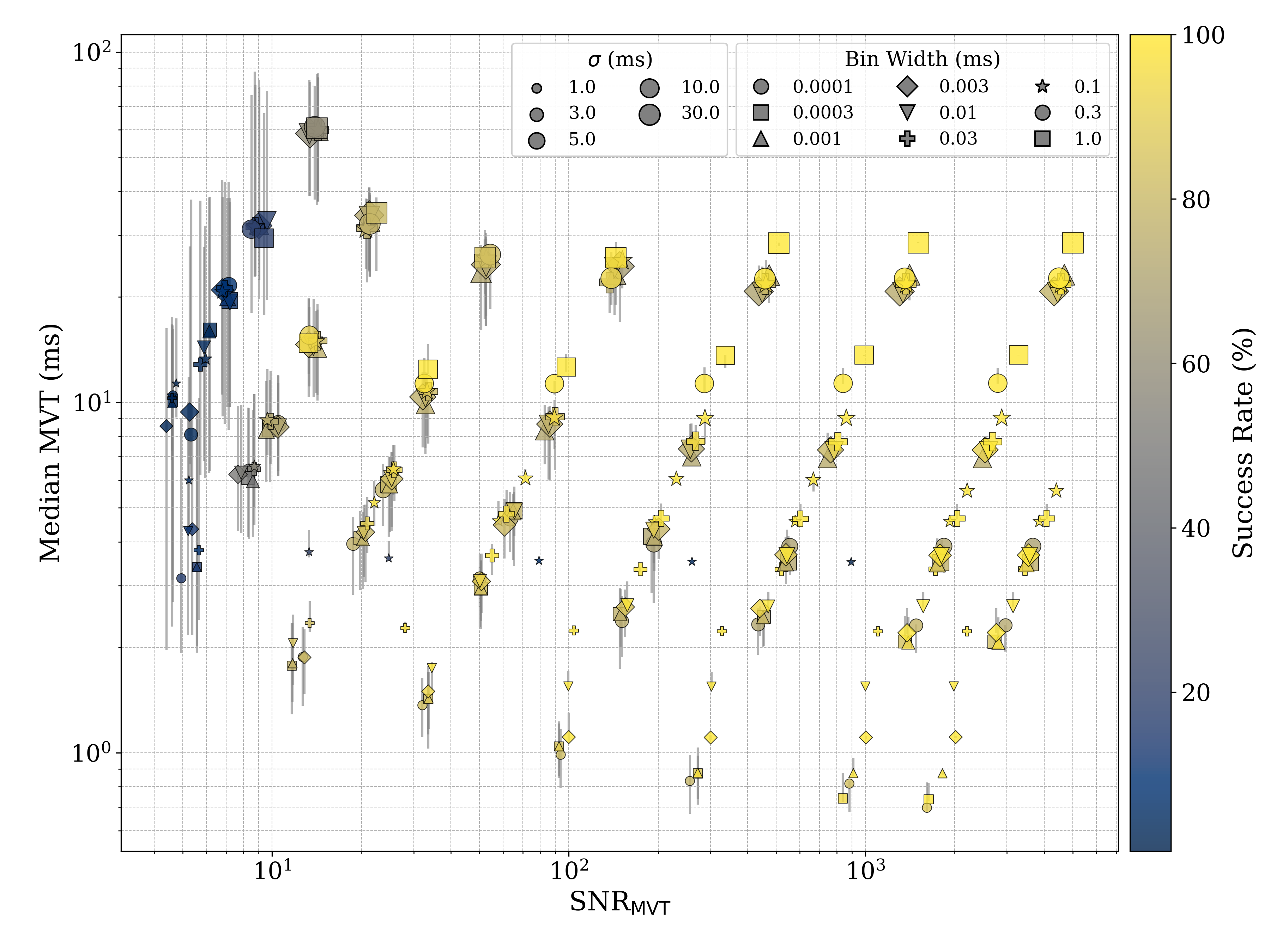}
    \caption{\small The median MVT as a function of the $\mathrm{SNR}_{\mathrm{MVT}}$ for the full suite of Gaussian pulse simulations. The size of each marker is proportional to the intrinsic pulse width ($\sigma$), while the color represents the measurement success rate. This plot provides a comprehensive visualization of the parameter space, confirming that reliable measurements (bright yellow points) are consistently achieved only at high $\mathrm{SNR}_{\mathrm{MVT}}$.}
    \label{appfig:gaussian_mvt_vs_snr_success}
\end{figure*}

Figure~\ref{appfig:gaussian_mvt_vs_snr_success} shows the median MVT as a function of $\mathrm{SNR}_{\mathrm{MVT}}$ for the full set of Gaussian pulse simulations. Each point represents one point in the input parameter grid, with marker size scaling with the intrinsic pulse width ($\sigma$) and color indicating the measurement success rate. This figure provides a global view of how recoverability depends on both intrinsic timescale and signal strength. As discussed in Section~\ref{sec:simulations}, reliable MVT measurements occur primarily in the high–$\mathrm{SNR}_{\mathrm{MVT}}$ regime, where the success fraction approaches unity. At lower $\mathrm{SNR}_{\mathrm{MVT}}$, the measurements become increasingly scattered and failure rates increase sharply, illustrating why $\mathrm{SNR}_{\mathrm{MVT}}$ must be accounted for when interpreting any single MVT determination.

\subsection{Norris Pulse Simulations}
\label{appsec:norris_sims}
\begin{figure*}[ht]
    \centering

        \includegraphics[width=.90\textwidth]{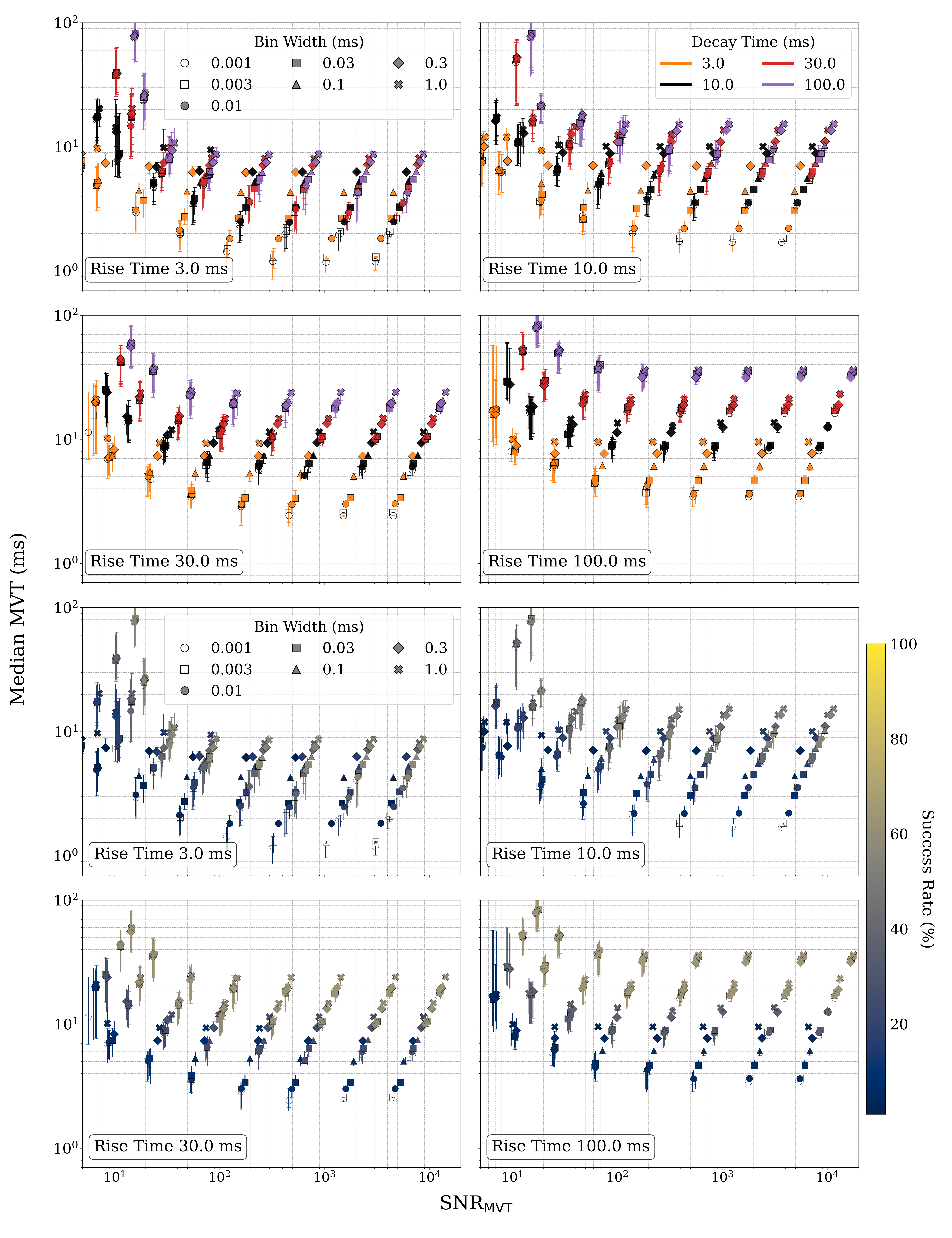}
        \caption{\small A comprehensive grid of MVT vs. $\mathrm{SNR}_{\mathrm{MVT}}$ results for Norris pulse simulations across a range of rise times (3, 10, 30, and 100 ms). \textbf{(Top two rows)} The data is colored by the pulse decay time, demonstrating that the timescale-dependent SNR threshold is a consistent feature regardless of the rise time. \textbf{(Bottom two rows)} The same data is colored by the measurement success rate.}
    \label{appfig:norris_mvt_snr_all}
\end{figure*}

Figure~\ref{appfig:norris_mvt_snr_all} shows the median MVT as a function of $\mathrm{SNR}_{\mathrm{MVT}}$ for the full set of Norris pulse simulations, grouped by rise time (3, 10, 30, and 100~ms). The top two rows are colored by decay time and demonstrate that the timescale–dependent SNR threshold persists across the full dynamic range of pulse asymmetry. The bottom two rows show the same data colored by measurement success rate, illustrating that high $\mathrm{SNR}_{\mathrm{MVT}}$ is a robust predictor of successful recovery regardless of the specific pulse morphology. As discussed in Section~\ref{sec:simulations}, these results confirm that the stability of an MVT measurement is controlled primarily by its location in the MVT–$\mathrm{SNR}_{\mathrm{MVT}}$ plane, not by the detailed shape of the pulse.

\subsection{Complex Light–Curve Simulations}
\label{appsec:complex_sims}
\begin{figure*}[htbp]
    \centering
    \rotatebox{90}{%
        \begin{minipage}{0.99\textheight}
            \centering
            \includegraphics[width=0.99\textheight]{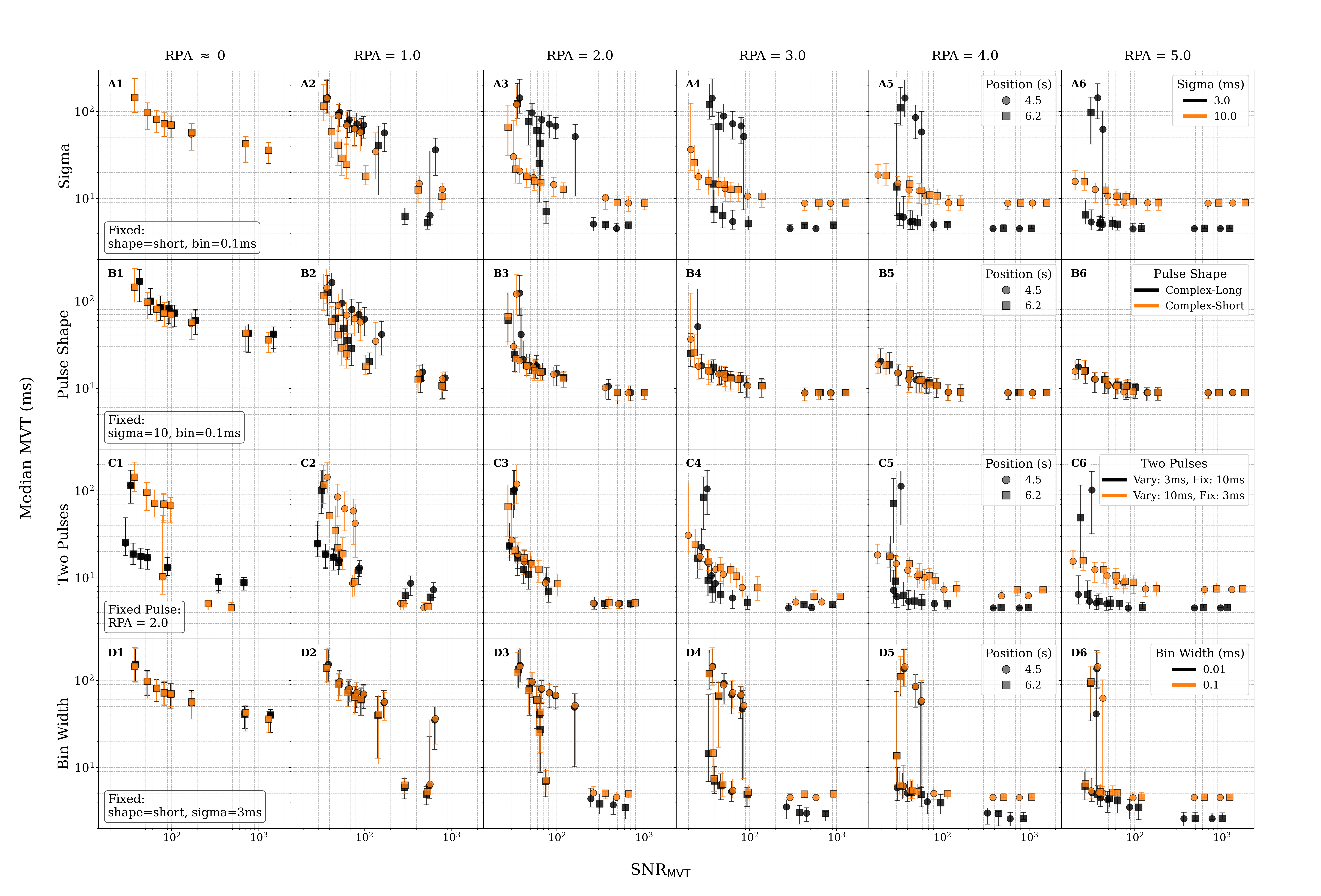}
            \caption{\footnotesize{A comprehensive overview of MVT parameter dependencies, showing the same data as Figure~\ref{fig:complex_mvt_vs_all_parameters} but plotted against $\mathrm{SNR}_{\mathrm{MVT}}$ instead of Overall Amplitude. }}
            \label{appfig:complex_mvt_vs_all_parameters_app}
        \end{minipage}
    }
\end{figure*}

Figure~\ref{appfig:complex_mvt_vs_all_parameters_app} shows the same simulation grid as Figure~\ref{fig:complex_mvt_vs_all_parameters}, but plotted in the MVT–$\mathrm{SNR}_{\mathrm{MVT}}$ plane rather than against overall amplitude. This representation directly illustrates the result discussed in Section~\ref{subsec:complex}: as the total signal significance increases, the MVT transitions from reflecting the broad template component to resolving the sharp injected substructure. This figure reinforces our conclusion that the apparent MVT in complex multi–component light curves is controlled not by the absolute signal amplitude but by the statistical strength of the smallest resolvable feature.

\subsection{MVT of Real GRBs Observed by \textit{Fermi}--GBM}
\label{app:source_intervals}

Table~\ref{apptab:fermi_gbm_details} provides a comprehensive breakdown of the MVT analysis results for our \textit{Fermi}-GBM GRB sample. It includes the results for different detector combinations and analysis bin widths, showing the Single MVT, the Median MVT from our Monte Carlo approach, the $\mathrm{SNR}_\mathrm{MVT}$, and the success rate of the measurements.

For several of the long–duration events in our sample (GRB~211211A, GRB~230307A, and GRB~250919A), the source intervals used in our MVT analysis do not span the full $T_{90}$ duration. This choice is primarily computational: generating hundreds of Monte Carlo realizations of very long light curves is time–intensive. However, this truncation does not affect measured MVT. The later phases of these bursts have much lower flux, and therefore contribute very little statistical weight to the MVT measurement, which is dominated by the bright, structured emission near peak intensity.

This approach is also consistent with our analyses of synthetic \textit{Complex–Long} and \textit{Complex–Short} simulations (Section~\ref{subsec:complex}), which showed that removing broad, low–amplitude components does not significantly alter the recovered MVT of the dominant emission episode. As a direct validation, we repeated the MVT analysis of GRB~230307A using its full source interval and obtained a statistically consistent result, confirming that the truncated intervals used in this work do not bias the final MVT measurements.

\begin{table*}[!htbp]
  \centering
  \caption{Detailed MVT Results and Analysis Parameters for the Fermi-GBM Sample.}
  \label{apptab:fermi_gbm_details}
  \resizebox{\textwidth}{!}{
  \begin{tabular}{l c c c c c c}
    \toprule
    Detector(s) & Bin Width (BW) & Single MVT & Median MVT & $\mathrm{SNR}_{\mathrm{MVT}}$ & Success & Failed \\
                & (ms)      & (ms)       & (ms)       &                             & (\%)    & (\%) \\
    \midrule
    \cmidrule(r){1-7}
    n2 & 0.1 & $7.7 \pm 1.6$ & $8.8_{-1.7}^{+2.0}$ & 134.7 & 88.3 & 11.7 \\
    n2, na, n1 & 0.1 & $4.2 \pm 1.3$ & $4.7_{-0.7}^{+0.9}$ & 121.9 & 58.7 & 41.3 \\
    n2, na, n1 & 0.01 & $2.6 \pm 1.4$ & $3.6_{-0.8}^{+1.6}$ & 113.5 & 46.3 & 53.7 \\
    n2, na, n1 & 0.003 & $2.5 \pm 1.3$ & $3.9_{-1.0}^{+1.0}$ & 116.4 & 42.7 & 57.3 \\
    All & 0.1 & $4.3 \pm 0.9$ & $4.4_{-0.6}^{+0.8}$ & 94.1 & 96.7 & 3.3 \\
    All & 0.01 & $2.5 \pm 0.9$ & $3.4_{-0.9}^{+0.9}$ & 85.0 & 58.3 & 41.7 \\
    All & 0.003 & $2.5 \pm 0.9$ & $3.1_{-0.7}^{+1.1}$ & 82.7 & 54.7 & 45.3 \\
    \multicolumn{7}{l}{\textbf{GRB 211211A} \quad Source: [-0.9, 15.0] s; Background: [-68.0, -18.0]; [69.7, 119.7] s; T$_{90}$: 34.3 s} \\
    
    \midrule
    \midrule
    \cmidrule(r){1-7}
    na, n6, n0, n7, n1, n9, nb & 0.1 & $4.2 \pm 0.6$ & $4.2_{-0.5}^{+0.4}$ & 135.4 & 98.0 & 2.0 \\
    na, n6, n0, n7, n1, n9, nb & 0.01 & $1.3 \pm 0.6$ & $1.9_{-0.4}^{+0.7}$ & 94.2 & 44.3 & 55.7 \\
    na, n6, n0, n7, n1, n9, nb & 0.003 & $1.3 \pm 0.5$ & $1.7_{-0.4}^{+0.7}$ & 90.9 & 36.3 & 63.7 \\
    All & 0.1 & $3.8 \pm 0.6$ & $3.9_{-0.3}^{+0.5}$ & 116.2 & 100 & 0 \\
    All & 0.01 & $1.5 \pm 0.5$ & $1.8_{-0.3}^{+0.4}$ & 82.1 & 62.3 & 37.7 \\
    All & 0.003 & $1.4 \pm 0.4$ & $1.8_{-0.3}^{+0.5}$ & 80.7 & 57.7 & 42.3 \\

    \multicolumn{7}{l}{ \hspace{19em} Source: [-1.64, 18.0] s} \\
    
    \cmidrule(r){1-7}
    na, n6, n0, n7, n1 & 0.1 & $4.3 \pm 0.8$ & $4.4_{-0.5}^{+0.7}$ & 137.0 & 93.0 & 7.0 \\
    \multicolumn{7}{l}{\textbf{GRB 230307A} \quad Source: [-1.6, 45.3] s; Background: [-68.8, -18.8]; [69.4, 119.4] s; \quad T$_{90}$: 34.56 s} \\
    \midrule
    \midrule
    \cmidrule(r){1-7}
    na, n1, nb, n4, n5, n0 & 1 & $383.7 <$ & $361.6_{-64.3}^{+85.9}$ & 7.5 & 31.7 & 68.3 \\
    na, n1, nb, n4, n5, n0 & 0.1 & $391.4 <$ & $344.6_{-141.7}^{+133.9}$ & 7.2 & 33.3 & 66.7 \\
    All & 1 & $256.7 \pm 74.6$ & $304.2_{-63.4}^{+113.6}$ & 7.0 & 65.3 & 34.7 \\
    All & 0.1 & $249.6 \pm 81.6$ & $332.2_{-85.5}^{+80.5}$ & 6.8 & 68.3 & 31.7 \\
    \multicolumn{7}{l}{\textbf{GRB 170817A} \quad Source: [-2.2, 4.3] s; Background: [-62.9, -12.9]; [14.9, 64.9] s; \quad T$_{90}$: 2.05 s} \\
    \midrule
    \midrule
    \cmidrule(r){1-7}
    n7, n6, nb, n8, n3, n9 & 0.1 & $8.9 \pm 2.1$ & $10.5_{-1.8}^{+2.0}$ & 24.2 & 82.0 & 18.0 \\
    n7, n6, nb, n8, n3, n9 & 0.01 & $14.8 <$ & $9.0_{-1.9}^{+2.8}$ & 24.1 & 62.7 & 37.3 \\
    n7, n6, nb, n8, n3, n9 & 0.003 & $6.6 <$ & $9.9_{-2.5}^{+2.1}$ & 24.1 & 66.0 & 34.0 \\
    All & 0.1 & $17.3 <$ & $12.4_{-3.9}^{+2.9}$ & 18.8 & 58.3 & 41.7 \\
    All & 0.01 & $8.8 <$ & $10.5_{-3.5}^{+4.1}$ & 18.9 & 41.0 & 59.0 \\
    All & 0.003 & $8.3 <$ & $10.3_{-3.1}^{+3.9}$ & 17.6 & 37.0 & 63.0 \\
    \multicolumn{7}{l}{\textbf{GRB 231115A} \quad Source: [-1.0, 2.0] s; Background: [-53.6, -3.6]; [7.2, 57.2] s; \quad T$_{90}$: 0.032 s} \\
    \midrule
    \midrule
    \cmidrule(r){1-7}
    n7, n8, n6, nb & 0.1 & $85.4 <$ & $29.3_{-4.2}^{+7.4}$ & 184.4 & 65.0 & 35.0 \\
    n7, n8, n6, nb & 0.01 & $60.8 <$ & $28.8_{-5.5}^{+10.7}$ & 180.5 & 62.3 & 37.7 \\
    All & 0.1 & $24.6 \pm 4.6$ & $29.7_{-8.4}^{+6.6}$ & 139.3 & 63.3 & 36.7 \\
    All & 0.01 & $23.6 \pm 5.1$ & $28.9_{-7.7}^{+8.8}$ & 137.9 & 62.0 & 38.0 \\
    \multicolumn{7}{l}{\textbf{GRB 250919A} \quad Source: [15.5, 41.0] s; Background: [-130.6, -20.6]; [238.1, 348.1] s; \quad T$_{90}$: 129.3 s} \\
    \midrule
    \bottomrule
    
  \end{tabular}
  }
\end{table*}




\end{document}